\documentclass[onecolumn]{emulateapj}
\usepackage{epsfig}
\usepackage{graphicx}
\usepackage{ifthen}
\usepackage{amssymb}
\usepackage{amsmath}
\usepackage{subfig}

\def\Deff{D_{\rm eff}}

\def\beq{\begin{equation}}
\def\eeq{\end{equation}}

\def\omega0{\Omega_{\rm m,0}}

\def\aj{{AJ}}% % Astronomical Journal
\def\araa{{ARA\&A}}% % Annual Review of Astron and Astrophys 
\def\apj{{ApJ}}% % Astrophysical Journal 
\def\apjs{{ApJS}}% % Astrophysical Journal, Supplement 
\def\aap{{A\&A}}% % Astronomy and Astrophysics 
\def\aaps{{A\&AS}}% % Astronomy and Astrophysics, Supplement 
\def\mnras{{MNRAS}}% % Monthly Notices of the RAS 
\shorttitle{Impacts of Source Properties on Strong Cluster Lensing}
\shortauthors{Gao, Jing, Mao, Li, \& Kong}

\begin{document}
\title{Impacts of Source Properties on Strong Lensing by Rich Galaxy Clusters}

\author{G. J. Gao$^{1,2}$, Y. P. Jing$^{1}$, 
S. Mao$^{3}$, 
G. L. Li$^{1,4}$,
X. Kong$^{5}$}

\affil{$^1$ Key Laboratory for Research in Galaxies and Cosmology,
Shanghai Astronomical Observatory, Chinese Academy of Sciences, 80
Nandan Road, Shanghai, 200030, China}

\affil{$^2$ 
Graduate School of the Chinese Academy of Sciences, Beijing,  100039, China}

\affil{$^3$ Jodrell Bank Centre for Astrophysics, School of Physics and Astronomy,
University of Manchester, Manchester M13 9PL, UK}

\affil{$^4$ Argelander-Institut f\"{u}r Astronomie, Auf dem H\"{u}gel
71, Bonn D-53121, Germany}

\affil{$^5$ Key Laboratory for Research in Galaxies and Cosmology, 
University of Science and Technology of China, \\ Chinese Academy of
Sciences, Hefei, 230026， China}

\email{gangjie.gao@gmail.com}

\begin{abstract}
We use a high-resolution $N$-body simulation to investigate
the influence of background galaxy properties, including redshift,
size, shape and clustering, on the efficiency of forming giant arcs by
gravitational lensing of rich galaxy clusters.  Two large sets of
ray-tracing simulations are carried out for 10 massive clusters at
two redshifts, i.e. $z_{\rm l} \sim 0.2$ and $0.3$. 
The virial mass ($M_{\rm vir}$) of the simulated lens clusters at $z\sim0.2$
ranges from $6.8\times10^{14} h^{-1} {M_{\odot}}$ to $1.1\times
10^{15} h^{-1} M_{\odot}$.  The information of background
galaxies brighter than 25 magnitude in the $I$-band 
is taken from Cosmological Evolution Survey (COSMOS) imaging data.
Around $1.7\times 10^5$ strong lensing realizations with these images as background galaxies
have been performed for each set. We find that the efficiency for forming giant arcs for $z_{\rm l}=0.2$ clusters
is broadly consistent with observations.  %\citep{Smith05}.
Our study on control source samples shows that the number of giant arcs 
is decreased by a factor of 1.05 and 1.61
when the COSMOS redshift distribution of galaxies is adopted,
compared to the cases where all the galaxies were assumed to be in a 
single source plane at $z_{\rm l}=1.0$ and $z_{\rm l}=1.5$, respectively.
% \citep{B98, Horesh05}.
We find that the efficiency of producing giant arcs by rich clusters is weakly
dependent on the source size and clustering. Our principal finding is that
%shape of background galaxies must be fully included in the modeling of
%giant arcs. 
a small proportion ($\sim 1/3$) of galaxies with elongated shapes 
(e.g. ellipticity $\epsilon=1-b/a>0.5$) can 
boost the number of giant arcs substantially.
Compared with recent studies where a uniform ellipticity
distribution from 0 to 0.5 is used for the sources, the adoption of directly 
observed shape distribution 
increases the number of giant arcs by a factor of $\sim2$. 
%In addition, we find that the source clustering has a negligible effect
%on producing giant arcs. 
Our results indicate that it is necessary to account for source
information and survey parameters (such as point-spread-function, seeing) to
make correct predictions of giant arcs and further to constrain the cosmological parameters.
\end{abstract}


\section{INTRODUCTION}

Giant arcs are spectacular examples of strong gravitational lensing found
in rich galaxy clusters. Background galaxies are stretched into 
long, thin arcs by the intense foreground gravitational field.
Hundreds of giant arcs 
have been found in both optically-selected and X-ray selected clusters
(e.g. \citealt{Luppino99, ZG03, Gladders03, Sand05}). 

Massive clusters are efficient producer of giant
arcs and the number of giant arcs is a good 
indicator of the abundance of massive clusters. 
The halo mass function is very sensitive to 
the cosmological parameters, especially at
the massive end.
Moreover, the internal
structures of massive clusters, such as substructures and ellipticity,
also depend on the cosmological parameters. They all affect the
lensing probability (optical depth, e.g., \citealt{Mene03a, Mene03b, Wambsganss04, Torri04,
Dalal04, Li05, OLS03, Oguri08, Hilbert07, Puchwein05, Puchwein09}) in various degree. 
Therefore, the observation of giant
arcs is a useful probe of the cosmological model, in particular the
matter power-spectrum normalization, $\sigma_8$ \citep{Li06b}, the matter density,
$\omega0$, and to a less extent the cosmological constant,
$\Omega_{\Lambda,0}$.

However, in order to use the observations of giant arcs to constrain
the cosmological models, one has to thoroughly understand how the lensing 
probability is affected by the distribution of background source galaxies, as well as
the intrinsic properties of lens population. It has
been shown that the lensing probability increases significantly with 
the increase of the source redshift, thus it is necessary to quantify
the redshift distribution of source galaxies in lensing studies
\citep{Wambsganss04}. The lensing efficiency of massive clusters does not
depend on the source size significantly \citep{Li05}, although not all
the real source information has been used, in particular their shapes.
%uncertainty because they did not use realistic information of source
%galaxies, such as their shapes and sizes.

\citet{Horesh05} first adopted realistic 
galaxy images with known photometric redshifts from the Hubble Deep Field (HDF) as background sources.
They selected clusters from a cosmological $N$-body simulation in a 
$\sim140 h^ {-1}{\rm Mpc}$ box at $z_{\rm l}\sim 0.2$. The mass
of the simulated lens clusters ($0.54$--$1.1\times10^{15} h^ {-1}{\rm  M_{\odot}}$) 
is similar to that of 10 X-ray--selected
 clusters \citep{Smith05}, the (mean) mass range of which is around 
$6.3\times10^{14} h^{-1} M_\odot \leq M_{200}\leq2.0\times10^{15} h^{-1} 
M_{\odot}$. Note that the mass is calculated based on the X-ray luminosity by using 
the $L_{\rm x}$-$M_{200}$ relation \citep{Reiprich02}.
They argued that the probability of producing giant arcs is $\sim 1$ per cluster after
observational effects are included, and emphasize that the number of
giant arcs produced by the simulated clusters in their adopted cosmology 
(with $\omega0=0.3, \Omega_{\Lambda,0}=0.7$ and 
the power-spectrum normalization of $\sigma_8=0.9$) 
is consistent with the
one observed in the massive clusters at redshift $0.171<z_c<0.255$. 

In this paper, we focus on the impact of background galaxy
properties, rather than the properties of the lens clusters,
on the efficiency of producing giant arcs. We use the
COSMOS data as direct input for the properties of background galaxies. We
investigate how the strong lensing probabilities are affected by the shape, size,
redshift distributions and clustering of background galaxies. Instead
of adopting all these source properties as a whole \citep{Horesh05},
we vary each of the source properties in turn
to gain a clear understanding about their individual impacts on strong
lensing statistics. The COSMOS data has a substantial advantage over
the HDF as it covers a much larger sky area (its sky coverage is 1.6 square degrees
vs. 5.3 square arcmins for the HDF) as used in \citet{Horesh05},
and thus is less affected by the cosmic variance.
While we confirm the effects of
source redshift and size distributions, we find that the shape distribution of
real galaxies has a dramatic effect on the lensing probability. With the
COSMOS images, the lensing probability is boosted by a factor of $2$
relative to that based on simple assumptions of the shape distribution
(e.g. random ellipticity between 0 and 0.5).
Our results indicate that the shape of
galaxies is an important factor in matching the theoretical
predictions and the observations of giant arcs. We will also show that
clustering of background galaxies has a negligible effect on the
lensing efficiency of producing giant arcs.

This paper is structured as follows. In \S2, we discuss COSMOS and background
source population.  In \S3, we discuss the simulation we use and our
lensing methodology.  In \S4, we present our main results and compare
with recent studies. We finish in \S5 with a short summary and a
brief discussion of the implications of our results.

\section{COSMOS and background source population}{\label{sec:srpop}}

COSMOS is a deep survey covering a $2$ square degrees of equatorial field 
containing over 2 million galaxies. It is an ideal sample
of background galaxies for our lensing simulations due to the excellent
resolution and large survey area (to reduce the cosmic variance). The
image is centered on RA=10:00:28.6 and DEC=+02:12:21.0 from the Hubble
Space Telescope Advanced Camera for Surveys (HST/ACS) \citep{Koekemoer07}.  Its 
calibration is relatively reliable due to the absence of atmospheric absorption. 
The $I$-band image data 
we use is the second public release of COSMOS observations. The camera
has two $2048\times4096$ CCD chips with a pixel scale of 
$0.05''$. All the drizzled image data  have been flux 
calibrated and astrometrically registered. The whole image in the $I$-band
has been cut into $575$ edge-overlapped tiles of
$4960 \times 4960$ pixels of CCD, which are generated by rotating with an angle of $\sim 10$ degrees to the tile edge and
embedded at the center of a rectangular box of $5600\times5600$ pixels. The 
left area in the box is empty of galaxies.

The galaxy images in the COSMOS field are identified with the
SExtractor program \citep{BA96}.  The SExtractor is a well developed 
software program for extracting
galaxy images from raw data and it produces a catalog of objects including 
properties such as their morphology, positions and
magnitudes. The SExtractor detection threshold
for all the ACS image data is set to be 1.5$\sigma$ which is about $25
\rm{mag/arcsec^2}$ in surface brightness.  Finally, a total of $\sim 3
\times 10^5$ galaxies are identified.
The output of SExtractor contains the relative positions of sources in
the tile and positions in the celestial equator coordinate system.
The relative position is useful for identifying which galaxies are strongly lensed in
the lensing process (see \S\ref{sec:tracing}).
The celestial coordinates of galaxies are matched with the positions in COSMOS
photometric catalog \citep{Mobasher07} of galaxies brighter than $I$-band 
AB magnitude of 25 to obtain the redshift information.
Only the galaxies successfully matched, thus having correct photometric redshifts, 
are added in our lensing simulations.
Using the COSMOS mask catalog, we exclude all  
star-like sources, galaxies close to star-like sources, and
those possibly contaminated by spikes due to the overflow of CCD
pixels. In particular, we use the same shape analysis program as in 
search for giant-arcs (see \S\ref{sec:tracing}) 
to identify very elongated sources with the major-axis to minor-axis ratio
$\geq10$ in all the tiles and exclude them 
to avoid ``giant-arc like'' artifact sources (e.g. imperfectly masked
spikes due to CCD pixel overflow and asteroid trails) or real sources which could have been lensed by foreground
objects in COSMOS field.
%, which could be incorrectly considered as giant
%arcs whether they are strongly lensed or not.
%
%{\bf (Q:What criterion is used? or you follow the mask catalog?
%where is the reference for the mask? A: The mask catalog is provided by 
%Prof. Kong. He told me he downloaded it from the COSMOS web site.)}. 

All the cleaned COSMOS tiles are added behind the lens clusters as 
source patches. The sky coverage of strong lensing simulation is 
the same as the original COSMOS field.
Due to the magnitude limit of the photometric catalog, the
density of matched galaxies with photometric redshift is about 500 per
tile of $5600\times5600$ pixels ($\sim 4.13\times4.13$ square arcmins) which contains 
$4096\times4096$ effective CCD pixels. Thus, the surface number density of the
extracted galaxies
is about 43 galaxies ${\rm arcmin^{-2}}$ in the CCD imaging area. 
Tiles of $5600 \times 5600$ pixels with white margin are placed randomly behind 
the lensing clusters in the lensing simulation, 
without specifically requiring that the lensing 
caustic area falls within the actual image.
This reduces the effective source number density
to $\sim 23$ galaxies ${\rm arcmin^{-2}}$ in our lensing analysis.

\begin{figure}
\epsscale{1.0}
\plotone{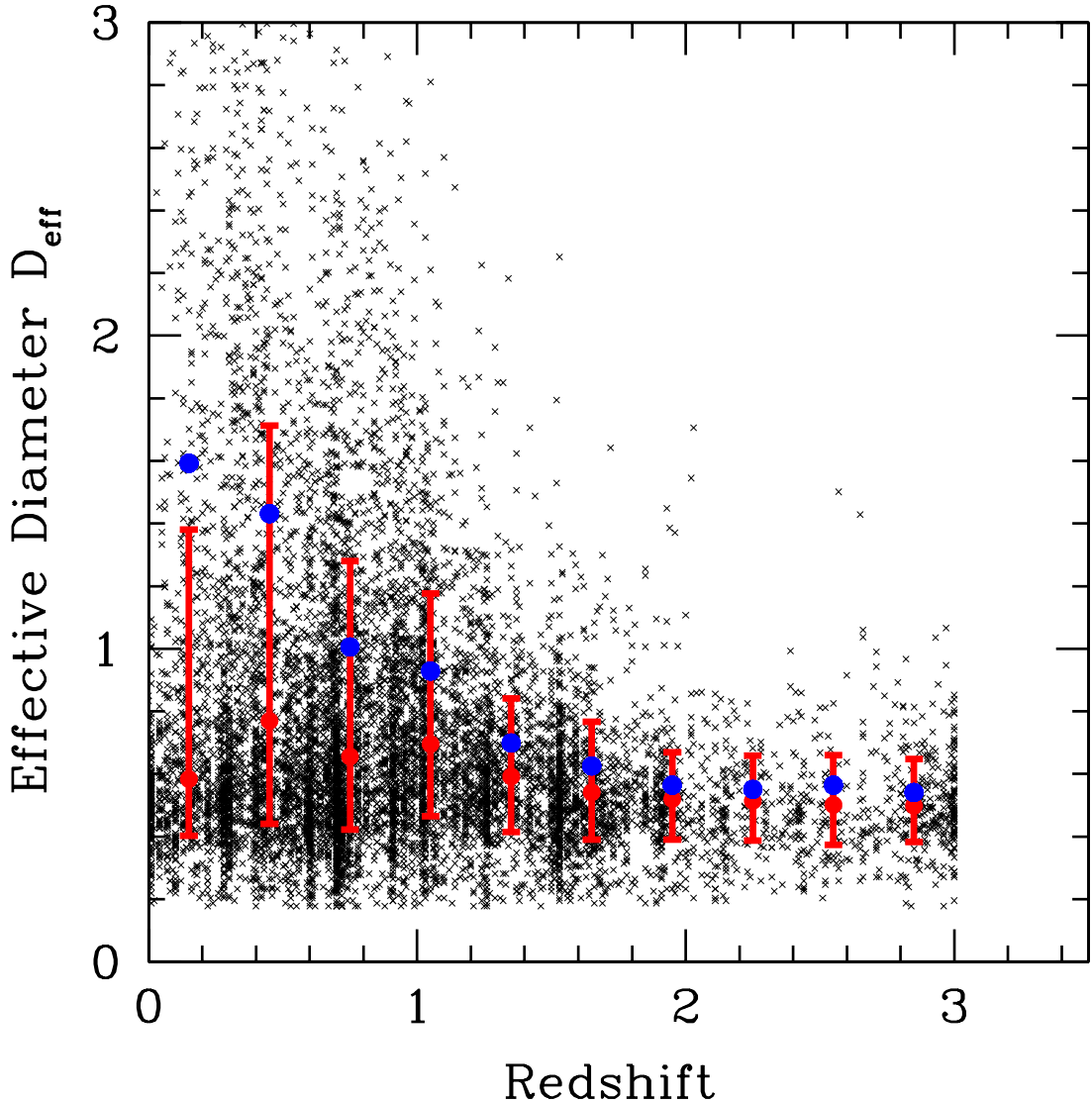} 
\caption{The scatter plot of size against redshift for galaxies randomly selected from 20 COSMOS image tiles with a bin width of $\Delta z=0.3$.
Each tile has $4096\times 4096$ effective pixels (with a pixel scale of 0.05 
arcsec by 0.05 arcsec) covering a field of view of 3.4 arcmins by 3.4 arcmins. 
The median (red filled circle) and the 68\% source size range (red bar) are
indicated in 10 bins of redshift. 
As can be seen, the source size distribution is almost independent of the redshift.
For comparison with \citet{Ferguson04}, the mean value (blue filled circle) of 
effective diameter in each redshift bin is also plotted. As our survey depth is 
much shallower, they actually agree well with each other.
}
\label{fig:sz}
\end{figure}

\begin{figure}
\epsscale{1.0}
\plotone{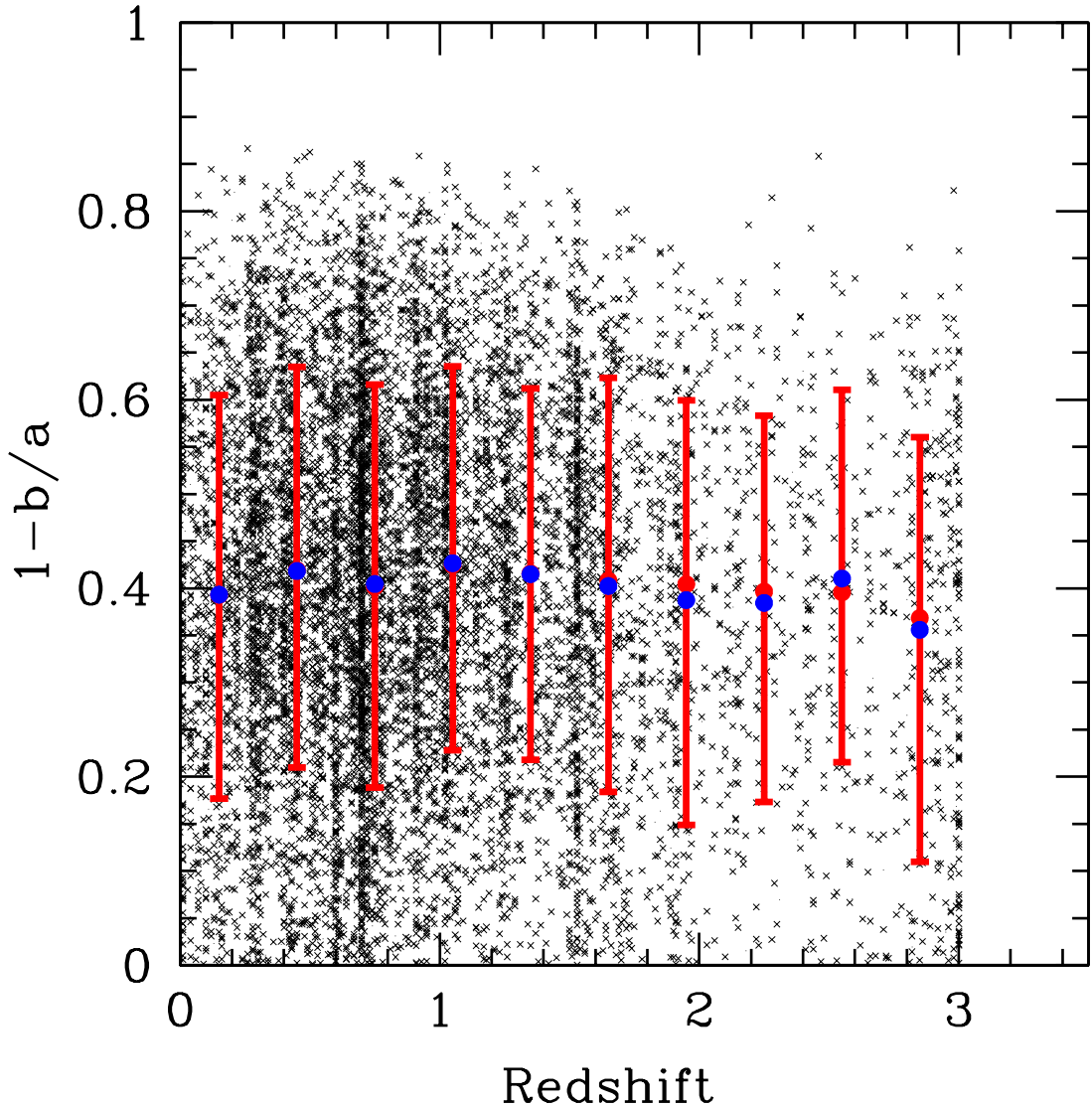} 
\caption{The scatter plot of the ellipticity, $\epsilon=1-{b/a}$, against redshift for 
galaxies randomly selected from 20 tiles with a bin width of $\Delta z=0.3$.
The median (red filled circle) and the 68\% ellipticity range (red bar) are indicated
in 10 bins of redshift.
As can be seen, the source ellipticity distribution is randomly distributed at 
all the redshifts.
The mean value (blue filled circle) of 
ellipticity in each redshift bin is also plotted for comparison. 
}
\label{fig:lwz}
\end{figure}

\begin{figure}
\epsscale{1.0}
\plotone{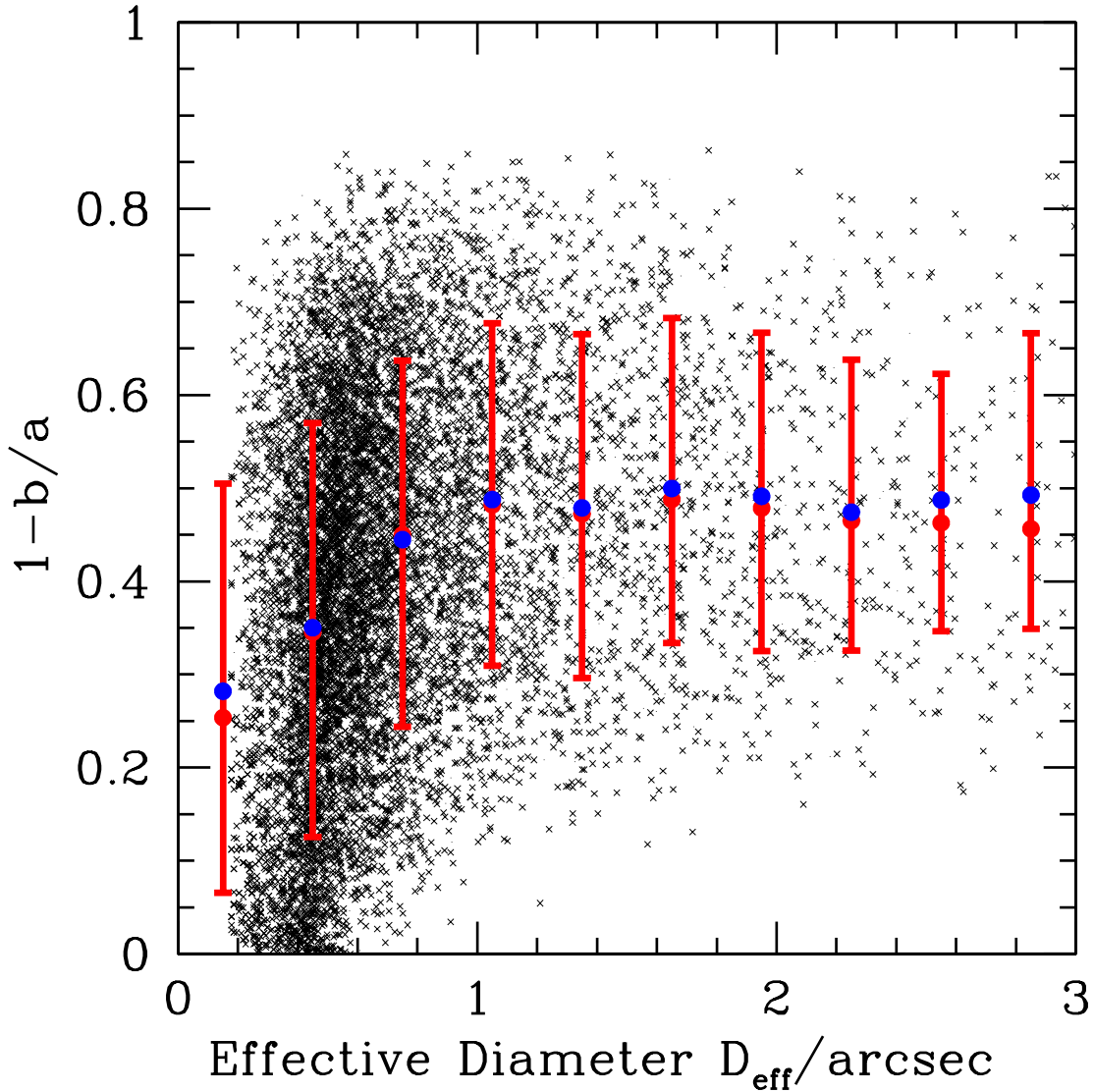} 
\caption{The scatter plot of the ellipticity, $\epsilon=1-{b/a}$, against
  size of galaxies in 20 randomly selected tiles with a bin width of 
$\Delta D_{\rm eff}=0.3$.
The median (red filled circle) and the 68\% ellipticity range (red bars) are indicated 
in 10 bins of effective diameter $D_{\rm eff}$.
The galaxies are more elliptical as the size increases for sources with 
the effective diameter smaller than $\sim 1.2''$, while the
median ellipticity keeps around $0.5$ for the left size range. 
The mean value (blue filled circle) of 
ellipticity in each effective diameter bin is also plotted for comparison.
}
\label{fig:slw}
\end{figure}

The scatter plots of source size (in pixels) vs. redshift, 
the ellipticity ($\epsilon$) vs. redshift and ellipticity vs. source size
with the median (as red filled circle) and the $68\%$ value range (as red bars) for
20 randomly selected source tiles are shown in Figs.~\ref{fig:sz}-\ref{fig:slw}.
The ellipticity $\epsilon$ is quantified by $1-b/a$, where $b$ and $a$ are semi-minor and
semi-major axes of a galaxy image. The axes are obtained by fitting the
image with an ellipse using the same method as that for quantifying the
shape of giant arcs (see \S\ref{sec:tracing}).
Fig.~\ref{fig:sz} shows that
the sources become smaller as the redshift increases from 0 to $\sim2$,
and then remain flat out to redshift $z=3.0$. However, the ellipticity of
galaxies is almost independent of redshift (see
Fig. \ref{fig:lwz}). The source size and ellipticity are somewhat
correlated (especially for the sources of the effective diameter 
smaller than $\sim1.2''$, 
as can be seen in Fig. \ref{fig:slw} -- larger sources
appear to be somewhat more elliptical. 

%The distributions of size, shape, and redshift of the
%galaxies are illustrated in Fig.~\ref{fig:sz}, \ref{fig:lwz} and
%\ref{fig:slw}. From the figures, it is clear that as the redshift
%increases, the sizes of sources appear to become smaller but their
%ellipticities remain virtually unchanged.

\begin{figure}
{
\epsscale{1.0}
\plotone{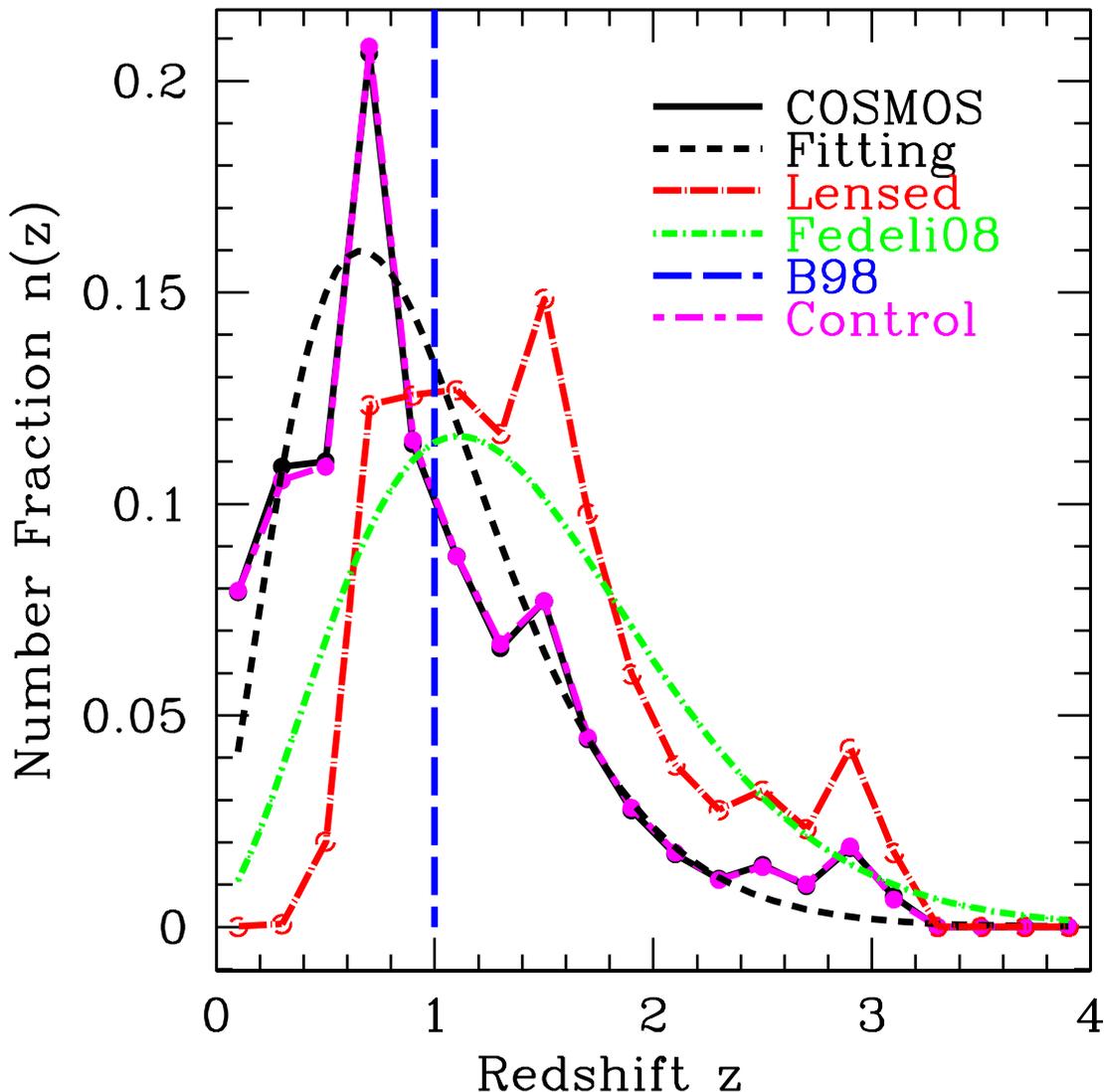} 
}
\caption{Comparison plot of redshift probability distributions between the 
subpopulation of original galaxies which have been 
strongly lensed to be arcs of L/W$\ge10$ ($z_{\rm l}=0.2$) and the parent population of COSMOS galaxies. 
The number fractions per redshift bin of width $\Delta z=0.2$ are plotted.
The lensed population is clearly shifted toward high-redshift end.
The fitting curve of the COSMOS redshift by Eq.~\ref{equation:fit} is indicated as the 
black dashed line.
	  The redshift distribution of the fiducial control sample consisting of 100 randomly selected tiles (used in Table~ \ref{table:cases}), which roughly follows the mean distribution, is plotted in magenta.
          For comparison, the redshift distributions used in \citet{B98} and
	  \citet{Fedeli08} are also shown.}  
\label{fig:zdist}
\end{figure}

The photometric redshift distribution of COSMOS background
galaxies is plotted (as the black solid line) in Fig.~\ref{fig:zdist}
for all the galaxies in $575$ tiles of COSMOS field. As can be seen
the redshift distribution peaks around $z=0.8$, but extends out
to $z=3.0$ with a median redshift of 0.78. The histogram is 
well fitted by the empirical formula from \citet{Smail95}:
\begin{equation}
p(z_{\rm s})=\frac{\beta}{z_0^3\Gamma(3/\beta)}{z_{\rm s}}^2
\,{\rm exp}\Big[-\left(\frac{z_{\rm s}}{z_0}\right)^\beta\Big],
\label{equation:fit}
\end{equation}
where $z_0\sim0.5$ and $\beta\sim1.2$; the best-fit line is 
indicated as the black dashed line in Fig. \ref{fig:zdist}.

With the photometric redshifts, we can construct an approximately three-dimensional
distribution of background galaxies in order to study the effect of
source redshift distribution on strong lensing efficiency.  A
series of source planes perpendicular to the line of sight is set
between redshift $z_{\rm l}\sim0.2$ ($\sim0.3$) to $z_{\rm s}=3.0$.  
The redshift separation between two adjacent
source planes $\delta z$ is set to $\delta {z/z}=0.1$ for
$1.5<z\leq3.0$ and to $\delta {z/z}=0.05$ for $0.2<z\leq1.5$.  
The galaxies are projected onto the planes according to which redshift bin they fall in. 
In each source plane, we mark all the galaxy pixels, which
are used to identify the lensed sources in subsequent ray-tracing.
There are 48 source planes in total for the ray-tracing program (see \S\ref{sec:tracing}).

\begin{figure}
{
\epsscale{1.0}
\plotone{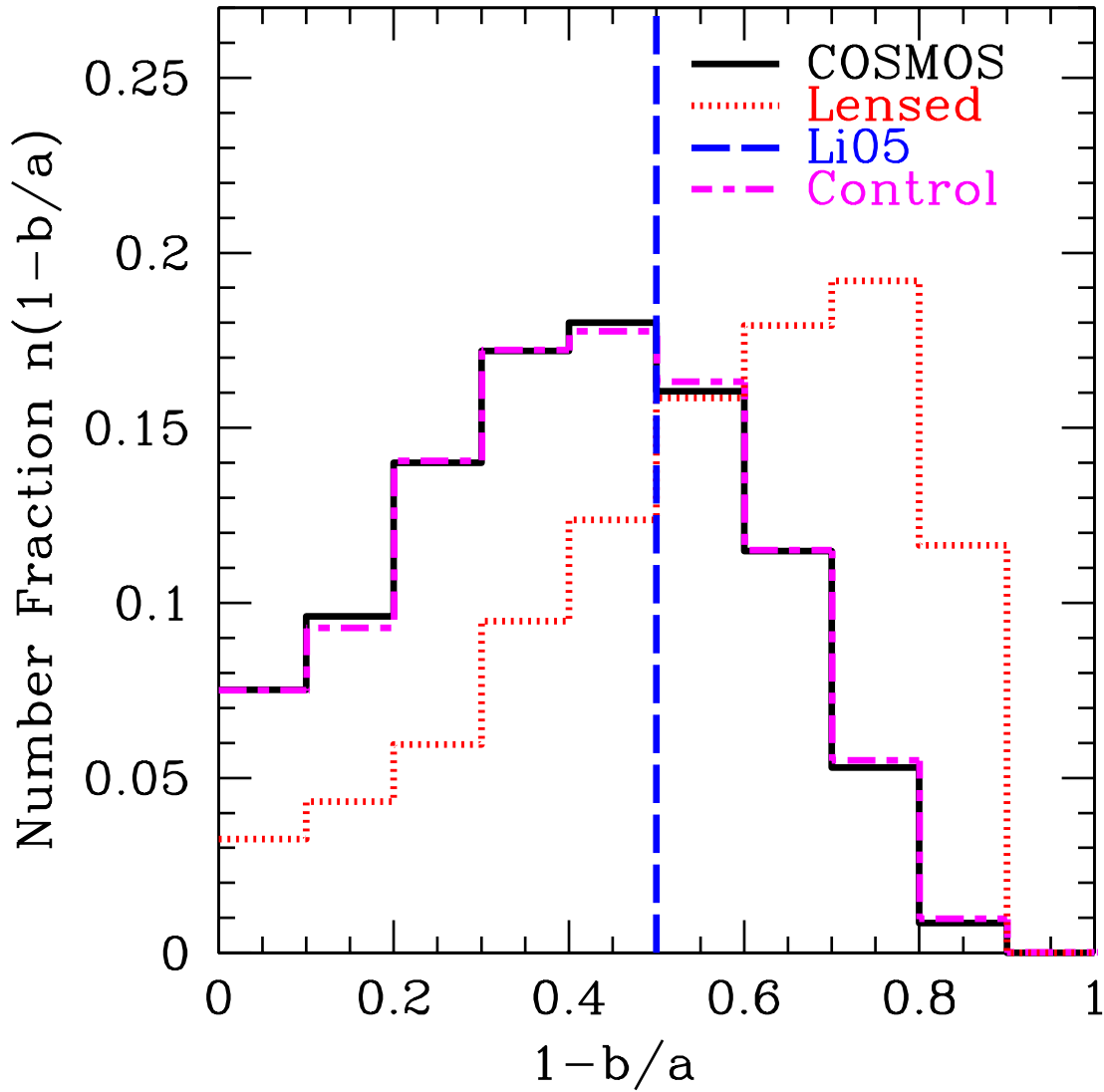} 
}
\caption{Comparison plot of shape probability distributions between the subpopulation of
	  original galaxies which have been strongly lensed to be arcs of L/W$\ge10$ ($z_{\rm l}=0.2$) and 
		    the parent population of COSMOS galaxies.  
The number fractions per ellipticity bin of width $\Delta\epsilon=0.1$ are plotted.
	  The lensed population clearly shows a bias
          towards high ellipticity compared with the original distribution.
	  The shape distribution of the fiducial control sample consisting of 100 randomly selected tiles (used in Table~ \ref{table:cases}) is plotted in magenta and agrees with COSMOS distribution very well. 
          For comparison, the distribution used in \citet{Li05} is also shown. 
}
\label{fig:lwdist}
\end{figure}

We will also study the impact of ellipticity and size distributions on
lensing cross-sections. 
The ellipticity distribution of the COSMOS galaxies 
are illustrated (as the black solid histogram) in Fig.~\ref{fig:lwdist}. 
The ellipticity $1-b/a$ distribution has a
median of 0.4, which is similar to that presented in \citet{Ferguson04}.
Compared to the studies
of giant arcs using a uniform distribution of the ellipticity with $0\leq1-b/a\leq0.5$ 
for background sources,  COSMOS clearly has an extended tail of highly flattened
galaxies, which will have a significant impact on the efficiency of
forming giant arcs (see \S\ref{sec:sourceshape}).

\begin{figure}
{
\epsscale{1.0}
\plotone{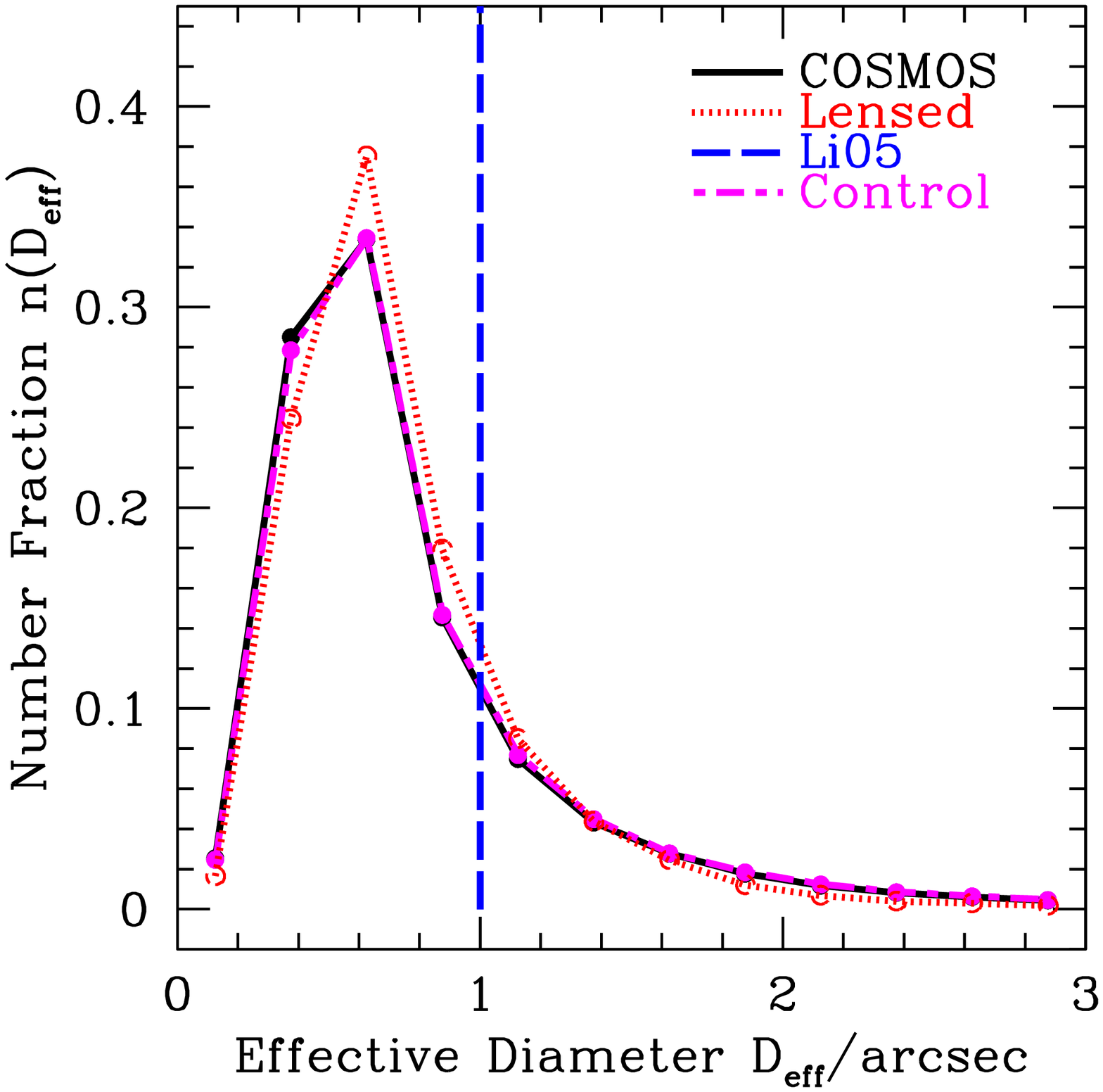} 

}
\caption{Comparison plot of size probability distributions between the subpopulation of
	  original galaxies which have been strongly lensed to be arcs of L/W$\ge10$ ($z_{\rm l}=0.2$) 
		    and parent population of COSMOS galaxies. 
The number fractions per effective diameter bin of width $\Delta
  D_{\rm eff}=0.25$ arcsec are plotted.
	  There is little difference between the two distributions, which
	  indicates they share the almost same population.
	  The size distribution of the fiducial control sample consisting of 100 randomly selected tiles (used in Table~ \ref{table:cases})
	  is plotted in magenta. There is no big difference from the mean in most 
		    $D_{\rm eff}$ range, except the places around the peak.
	  For comparison, the distribution used in \citet{Li05} is also indicated.
}
\label{fig:sdist}
\end{figure}

The size of each galaxy is quantified by an effective diameter $D_{\rm eff}$ 
defined as the diameter of a circular-shaped galaxy with the same number of the image
pixels. The size distribution of galaxies in these 
tiles is shown (as the black solid line) in Fig. \ref{fig:sdist}. The size distribution of galaxies has
a peak at $D_{\rm eff} \sim 0.75''$, but ranges from effective diameter $D_{\rm eff}<0.2''$
to $D_{\rm eff}>2''$. The median diameter of background galaxies is about $0.6''$,
   smaller than that assumed to be $D_{\rm eff}=1''$ \citep{Li05} since a  
   relatively deeper galaxy survey is adopted in this work.

%%%%%%% f1:sz.eps%%%%%%
%%%%%%%%%%%%%%%%%%%%%%%%%%%

%%%%%%% f2:lwz.eps%%%%%%
%%%%%%%%%%%%%%%%%%%%%%%%%%%

%%%%%%% f3:slw.eps%%%%%%
%%%%%%%%%%%%%%%%%%%%%%%%%%%

%%%%%%% f4:zdist.eps%%%%%%

%%%%%%%%%%%%%%%%%%%%%%%%%%%

%%%%%%%%%%%%%%%%%%%%%%%%%%%
%%%%%%% f5:lwdist.eps%%%%%%
%%%%%%%%%%%%%%%%%%%%%%%%%%%

%%%%%%% f6:sdist.eps%%%%%%

\section{Numerical Simulation and Lensing Methodology}

\subsection{Simulated Clusters}

The lens clusters are selected from a $N$-body simulation generated with a upgraded 
version of vectorized-parallel ${\rm P^3M}$ code of \cite{JS02} \citep{JSM07}. 
The underlying cosmological model is a $\Lambda$CDM ($\Omega_{\rm m,0}=0.268$,
$\Omega_{\Lambda,0}=0.732$, $n=1$) model. The simulation uses
$1024^3$ dark matter particles in a box with a side length of 600${h}^{-1} \rm {Mpc}$
(comoving). The mass resolution is $1.5\times10^{10}h^{-1}M_\odot$, 
thus massive clusters are reasonably resolved (with more than $4\times10^4$ particles in our samples). 
In this simulation (refer to \cite{JSM07} for more details), 
the Hubble constant $h$ in unit of $100\,{\rm km\ s^{-1} \,
Mpc^{-1}}$ is taken to be 0.71, while the amplitude of the
linear density power spectrum, $\sigma_8$, is 0.85, which is obtained
from CMBFAST \citep{Seljak96} directly. Particle pairwise interactions
are softened on scales smaller than 30 $h^{-1}\rm{kpc}$.  The
clusters are identified by the Friends-Of-Friends (FOF) method with
a linking length of 0.2 times the mean particle separation, and the
cluster mass $M_{\rm vir}$ is defined as the mass enclosed within the
virial radius according to the spherical collapse model \citep{KS96,
BN98, JS02}.

We primarily use a sample of 10 massive clusters at redshift $ z\sim0.2$
to perform ray-tracing simulations. This is to facilitate comparisons with 
observations of the X-ray selected cluster sample \citep{Smith05}.
The mass of the massive clusters at redshift $z\sim0.2$ ranges from
$6.8\times10^{14}  h^{-1} M_\odot$ to $1.1\times10^{15} h^{-1}
M_{\odot}$, roughly consistent with the observed range of mass
($6.3\times10^{14}\leq M_{200}\leq2.0\times10^{15} h^{-1} M_{\odot}$ of 
clusters at redshift $0.171<z_c<0.255$).\footnote{Notice that 
$M_{200}$ is defined as the mass enclosed within in $r_{200}$,
the radius within which the average density is 
equal to 200 times the critical cosmological density. The difference between $M_{200}$ and
$M_{\rm vir}$ are within 10\% for 8 of 10 clusters. 
It is within 30\% for the other two clusters.}
We shall also use 10 most massive clusters at $z \sim 0.3$ for 
lensing efficiency comparisons.

\subsection{Ray-tracing and Mapping}{\label{sec:tracing}}

We use the thin lens plane approximation for the lensing calculation;
the line of sight contributions may be important \citep{Wambsganss04, Puchwein09}. 
The surface density is calculated on a mesh of $1024\times 1024$
(coarse) grids
centered at the cluster with a size of 2 times the virial radii.
%the three coordinate axes of the simulation box.  
Our smoothing method, similar to \citet{Li06a}, uses the smoothed particle
hydrodynamics kernel \citep{Monaghan92} to distribute the mass of
particles enclosed inside the virial radius of the lens onto 3-D
grids and then the surface density is obtained by integrating along
the line of sight. The dimensionless surface density ($\kappa$) map is obtained
by dividing the surface density by the critical value,
\begin{equation}
\Sigma_{\rm cr}=\frac{c^2}{4\pi G}\frac{D_{\rm s}}{D_{\rm l} D_{\rm{ls}}}, 
\end{equation}
where $D_{\rm s}$, $D_{\rm l}$, and $D_{\rm {ls}}$
are the angular diameter distances between the source and the observer, the lens and the
observer, and the source and the lens.
The lensing potential on the grids of the coarse mesh,
$\phi$, is calculated by the Fast Fourier Transform (FFT) method from the dimensionless surface
density map since they are related by $\nabla^2 \phi = 2\kappa$. The
shear ($\gamma$) is then obtained by the second-order derivatives of the
potential, and the (signed) magnification is given by
$\mu^{-1}=(1-\kappa)^2-\gamma^2$.

To better determine the geometry of an image and to perform an efficient lensing simulation,
we refine the cell size of the
central high magnification area with $|{\mu}|\geq2.5$ in the lens plane up to
$0.05''$, identical to the intrinsic observational resolution in the source planes.  
High resolution is necessary for resolving the lensed image well, especially in the 
direction of minor axis, which is much more sensitive to the resolution.
The lensing potential on the refined mesh is obtained by cubic spline
interpolations of the neighboring $14\times14$ coarse grids in the 
$1024\times1024$ mesh.  Once we obtain
the lensing potential in the whole fine mesh area of interests, the
deflection angle on each grid point can be calculated from the
derivative of the lensing potential, $\vec{\alpha}=\nabla\phi$. The lens mapping
from the lens plane to each source plane at different redshifts can be
constructed using the lens equation (i.e. ray-tracing), $\vec{y}=\vec{x}-\vec{\alpha}$, where
$\vec{y}$, $\vec{x}$ are dimensionless source position in source plane and image position
in lens plane.
The ray-tracing procedure is repeated for each of the
$\sim50$ source planes, as we discussed in \S\ref{sec:srpop}.

We identify the image pixels in the lens plane 
from the mapping between the lens plane and the source planes.
For each projection of a lens, we go through all the refined grid points one by one and 
check
whether there is a corresponding point in the source planes falling into a galaxy. 
If so, we identify this grid point in the image plane as a lensed image pixel of that source galaxy;
if not, it is a trivial grid.
All the highly magnified lensed image pixels (e.g., $|{\mu}|\geq2.5$) 
can be located.  Finally, we can obtain the whole image(s) of a background
galaxy by collecting all the image pixels of the same source using 
Friends-of-Friends (FOF) method.

For source slices at different redshifts, the high-magnification areas will be 
mapped into areas of different sizes in the sky. 
We choose a rectangular conjunctive angular area inclusive of every caustic area 
and put this angular box randomly in the source tile to select different background 
galaxies for the ray-tracing simulation.
The box size is usually on the level of a few square arcmins (depending on
projection and also mass profile).
Since each added source tile is about $4.7' \times4.7'$, 
around 10 random lensing realizations are needed to cover a tile effectively. 
To avoid boundary effects, the source tile is periodically expanded in both
dimensions which are then randomly sampled.
%
%We choose the lensing area by shifting the maximal rectangular box in the
%periodically two-dimensionally expanded source tile by 10 times.  
%
%The source tiles are added one by one to do the ray-tracing simulation. 

The same process is carried out for all the three orthogonal 
projections of the massive clusters at the expected redshifts. 
10 ray-tracing simulations are carried out with all source galaxy samples in
the COSMOS field for each of 3 orthogonal projections.
The lensing images found are stored and subject to more detailed
analysis to find their length-to-width (L/W) ratios, sizes,
the redshifts.
Of particular importance is the L/W ratio. We determine this quantity
following \cite{B98} and \cite{Li05}. Briefly, we first identify the center of
the image pixels for a given source ($\vec{x}_c$), and then locate
the point $\vec{x}_1$ that is the furthest from the centre, and finally the
point $\vec{x}_2$ that is the furthest from $\vec{x}_1$. We then fit a circular arc
that passes through these three points ($\vec{x}_1, \vec{x}_2,
\vec{x}_{c}$). The half length of
the arc is taken to be the semi-major axis length of the ellipse, $a$, and
the length of the semi-minor axis is taken to be $b=S_{\rm image}/(\pi
a)$, where $S_{\rm image}$ is the area covered by the image.
The length-to-width ratio is then simply $a/b$. As mentioned before, the same
procedure is used to determine the ellipticity, $1-b/a$, for unlensed galaxies.

%The redshifts of the lensed images corresponding to their sources
%are also recorded.
%Because when the pixels of a galaxy with known redshift are lensed,
%we consider the corresponding images hold the same redshift information.

\subsection{Simulating test images}

\begin{figure*}
\begin{center}
{
\subfloat[outside]{\includegraphics[width=0.32\textwidth]{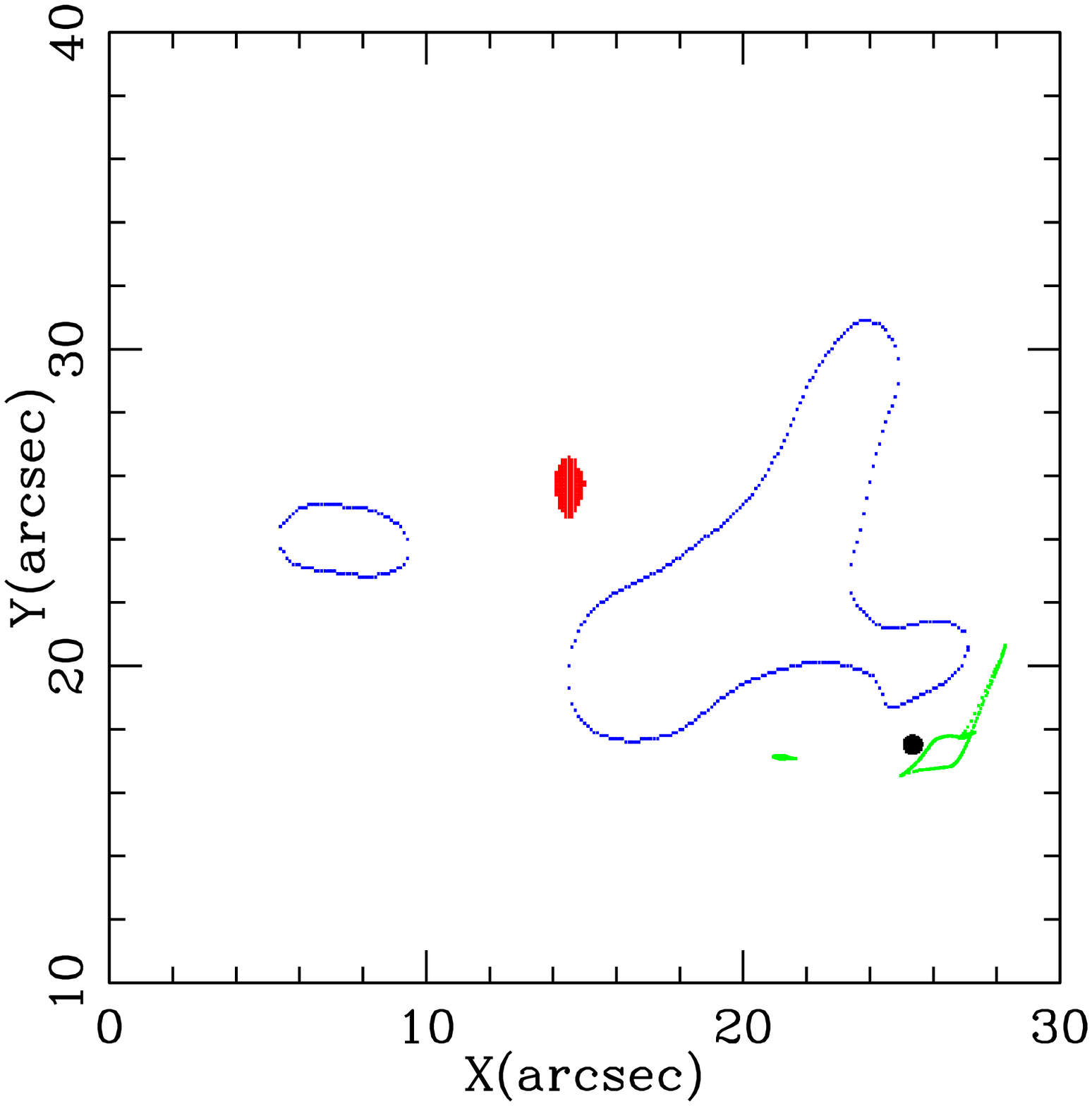}}
\subfloat[across]{\includegraphics[width=0.32\textwidth]{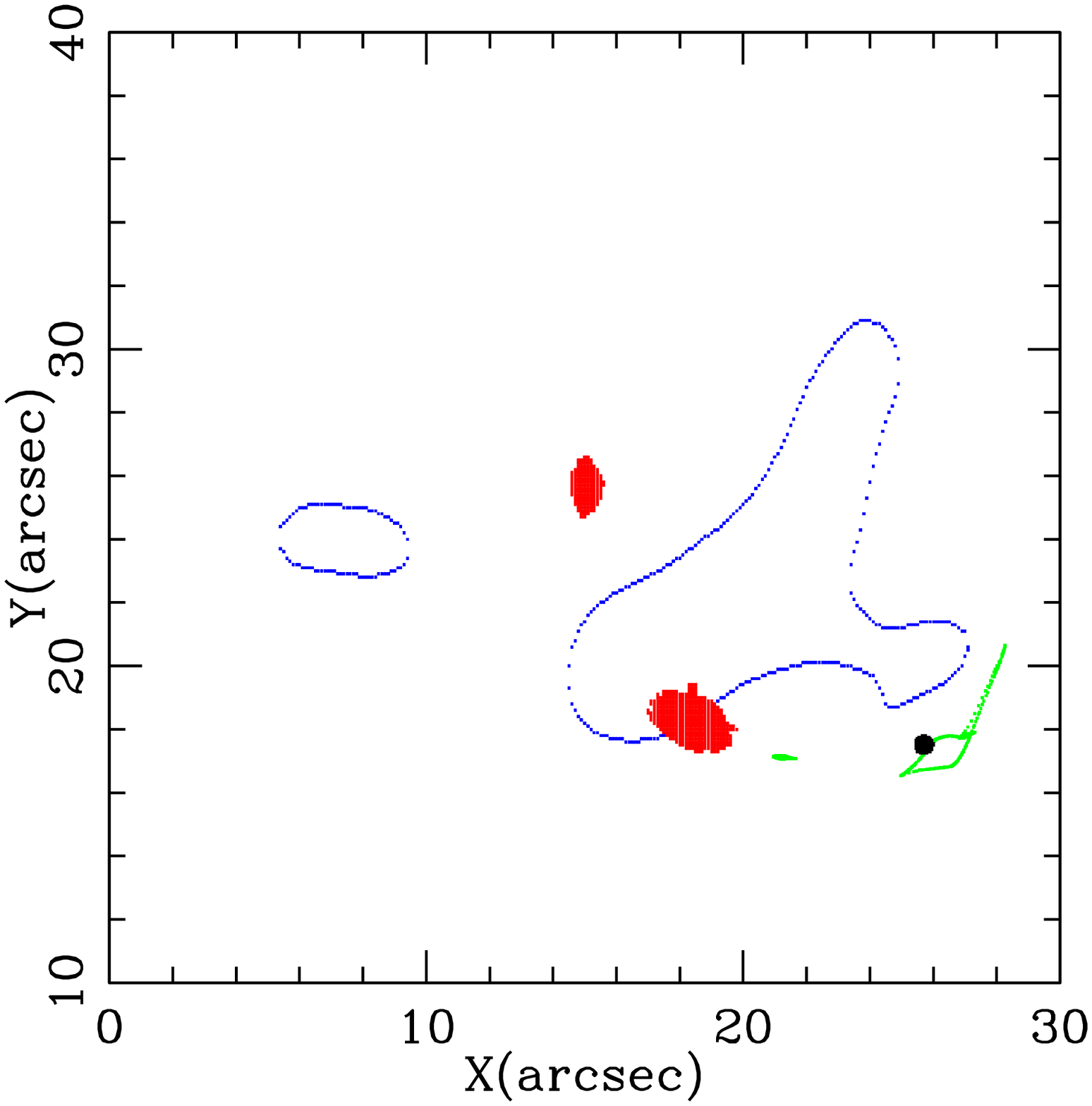}}
\subfloat[inside]{\includegraphics[width=0.32\textwidth]{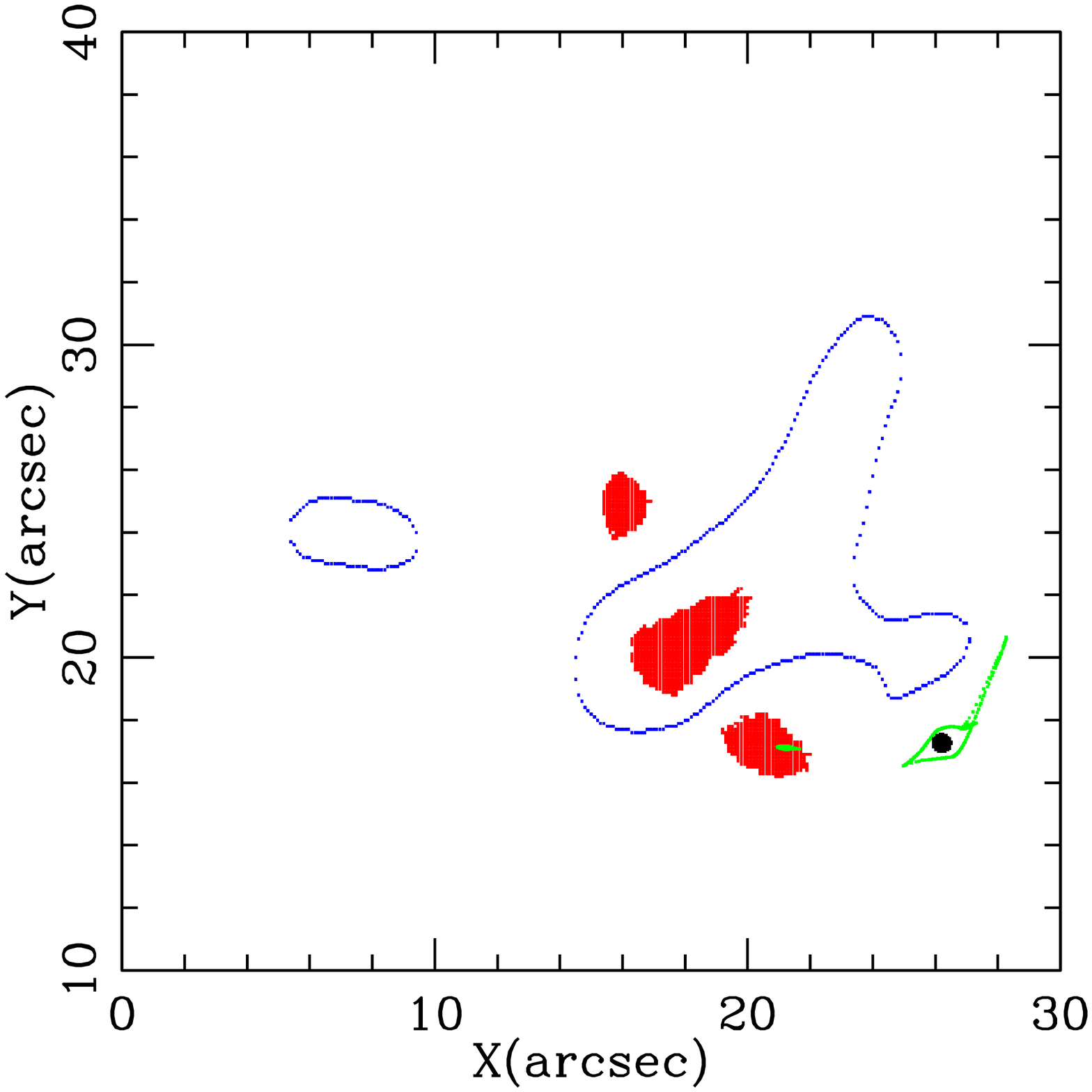}}
}
\caption {Simulated images for a mocked circular galaxy located outside, on and inside the tangential
	  caustic of a merging galaxy cluster, which is selected to illustrate for 
easy identification.
	    The black dots indicate the sources, while the red ones are
	 the lensed images. The green and blue dotted lines indicate the
	 caustics and critical curves respectively. As can be seen, the lens produces 
	 single, double and triple strongly lensed images.}
\label{fig:mock}
\end{center}
\end{figure*}

\begin{figure}
{
\epsscale{1.0}
\plotone{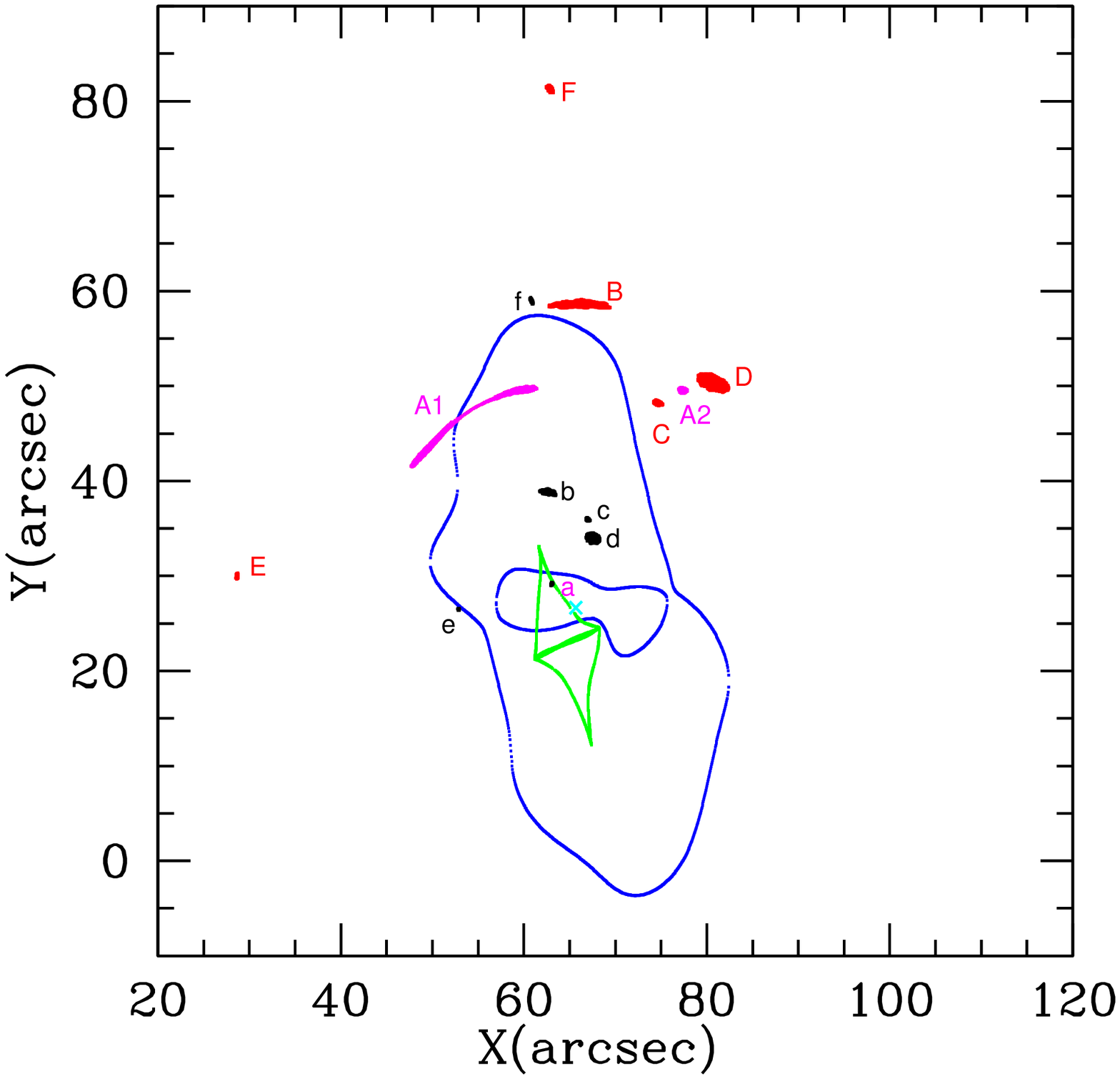} 
}
\caption{ Simulated images by a massive lens cluster ($z_{\rm
l}=0.3$) for galaxies in a COSMOS tile. The sources
and their lensed images are shown in black and red respectively.
The other unlensed galaxies in the tile are not plotted together for a clear view .
The blue and green curves are critical lines and their corresponding caustics for 
sources at redshift $z_{\rm s}=3.0$,
while the cyan cross indicates the center of lens.  A strongly stretched
arc ``A1'' is clearly seen corresponding to the source ``a''; an additional
image ``A2'' is highlighted (in magenta) for the same source.
}
\label{fig:simulatedImage}
\end{figure}

We carried out a test by putting a source around the
tangential caustic curve (as the green dotted line) of a massive simulated cluster. 
The simulated images (shown in red) are illustrated in Fig.~\ref{fig:mock} together
with the source (shown in black) positions. 
As expected, it produces one, two and three images when the mocked
source is outside, on and inside the caustic curve.
Fig.~\ref{fig:simulatedImage} shows a simulated
image of a massive lens cluster of a realization based on a real source tile. Clearly, the
sources (shown in black) have been systematically stretched around the cluster center. 
In particular, there is one giant arc at redshift $z=2.52$ (labeled as ``A1'') with $L/W\sim22$ among the lensed
images produced from source ``a'', which has an additional image labeled as ``A2''. 
Notice that the
positions of the images and background galaxies in different planes are transformed
into the same coordinate system centered at the cluster centre.
%Because of our high mesh resolution on the lens plane, we can have very detailed
%information of the edge of the lensing images.

%%%% f7=op_***.eps%%%%%%
%%%%%%%%%%%%%%%%%%%%%%%%%%%

%%%% f8=simge.eps%%%%%%
%%%%%%%%%%%%%%%%%%%%%%%%%%%

%\input{res}
\section{RESULTS}{\label{sec:results}}

Many recent studies have shown that the efficiency of forming giant
arcs depends on the background source population, in particular, their
sizes and redshifts \citep[e.g.,][]{Wambsganss04, Li05, Horesh05}.
Since different giant arc surveys
have different selection effects, the goal of this work is not only to
compare our strong lensing efficiency with observations, but also to focus on 
using realistic, empirical information of background sources, and then
to study how the lensing efficiency changes as we vary the assumptions.

For this purpose, we construct 7 comparison source samples in the control group 
which will be added in the ray-tracing simulations.
All the samples are based on 100 randomly selected COSMOS tiles. The
mean number density of this control sample is around 500 galaxies per
tile, which is the same as the mean
density from the whole field. The sample with COSMOS background sources
is our fiducial case (Case 1).
This sample also provides the basis on which we generate
mocked samples with partially or fully artificial source properties 
for Cases from 2 to 7.  The artificial sources are mocked tile by tile.
The lensing simulations are performed for 3 projections of 5 massive clusters. 
For effectively sampling the source area, 10 realizations are carried out for each 
projection in each source tile. There are 15000 simulations in total for each case.
The resulting arc statistics for this control group (e.g. six mocked 
samples together with the COSMOS sample) are listed in Table~\ref{table:cases}. 
The lensing efficiency changes as different source properties are switched. 
We find source properties affects the lensing efficiency in various degrees by 
comparing the numbers of giant arcs between different cases, especially the source
ellipticity and redshift. Furthermore, we also find that the boosting factor
due to the source ellipticity on lensing efficiency depends on the source redshift.
The details are listed below.

\subsection{Lensing efficiency of COSMOS sources}

\begin{figure}
{
\epsscale{1.0}
\plotone{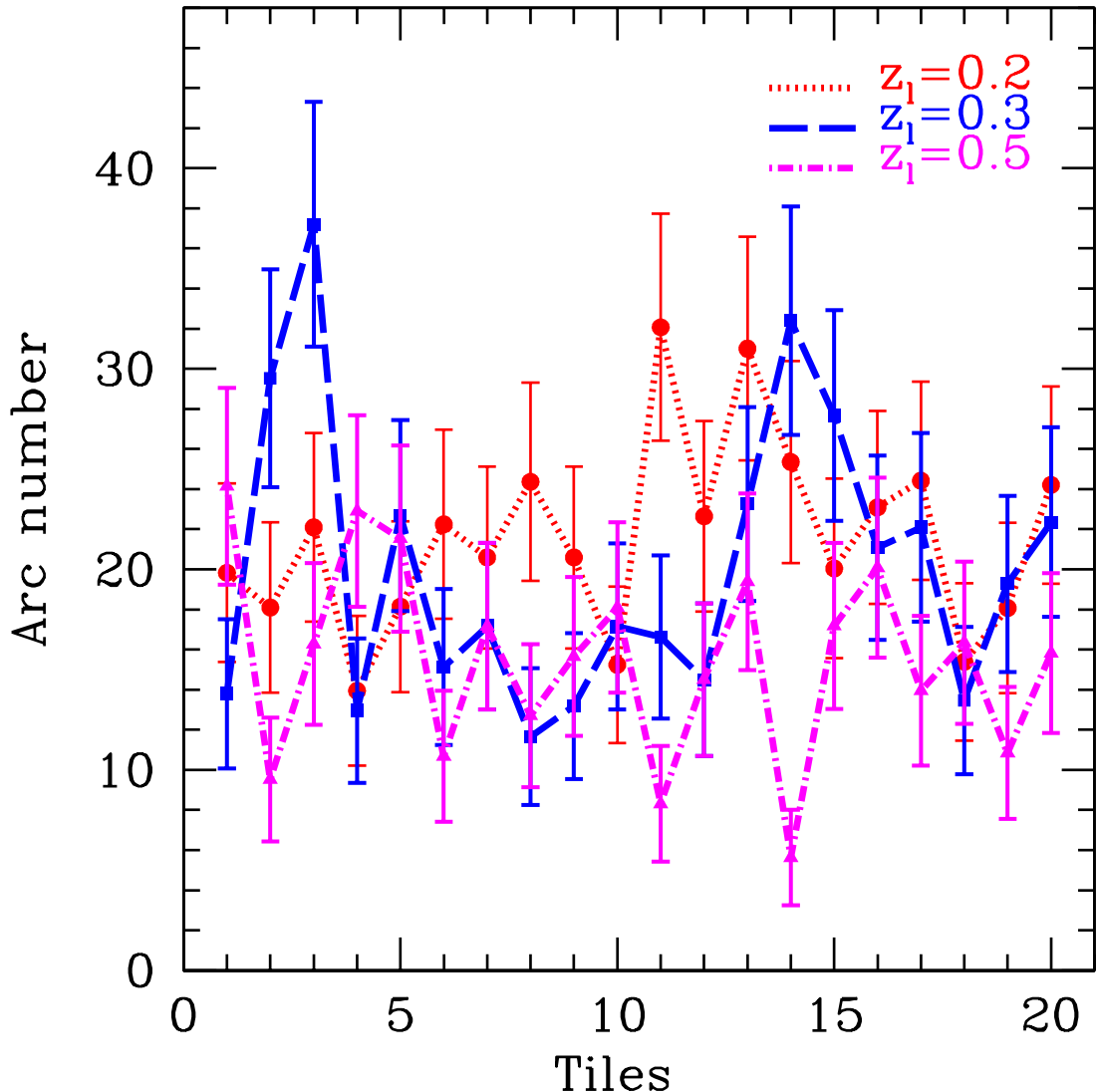} 
}
\caption{The giant-arc statistics for three test runs at different lens
        redshifts, e.g., $z_{\rm l}=0.2$, 0.3 and 0.5 are illustrated. 
		    The 20 test source tiles are randomly selected.  
		    For each source tile,  all the giant arcs produced in 300
realizations of 10 massive clusters are counted (see Table \ref{table:simu}). As can
be seen, the fluctuation is large for a single source tile in the mock survey due to the
limited realizations. The lensing efficiency decreases when the lens redshift ranges from
0.2 to 0.5, which is consistent with the result in \S\ref{sec:le}.
}
\label{fig:com_steps}
\end{figure}

Using the empirical COSMOS background source information, and the
method outlined in the last section, we implement a mock lensing survey for each 
projection of the massive clusters at redshift $z_{\rm l}=0.2$ 
and $z_{\rm l}=0.3$ for all of 575 tiles from COSMOS (Table~\ref{table:simu}).
We adopt the definition of the $L/W$ ratio of \citet{Li05}. In total,
12816
giant arcs of $L/W\geq10$ are found out of 172500 cases 
at $z_{\rm l}=0.2$, while 9704 giant arcs for clusters at $z_{\rm l}=0.3$. Therefore, 
the lensing efficiency is about 0.0743 giant arcs per cluster at $z_{\rm l}=0.2$ and 0.0563 giant arcs per cluster at $z_{\rm l}=0.3$.
We illustrate the lens redshift ($z_{\rm l}$) dependence and the fluctuation of
number counting in Fig.~\ref{fig:com_steps} for 20 randomly selected tiles.
All the giant arcs ($L/W \geq 10$) produced by the 10 massive lens clusters
from the galaxies in each tile are taken into account. The lensing
probability appears to decrease from $z_{\rm l}=0.2$ to $z_{\rm l}=0.5$ in our case, although the
fluctuation is relatively large mainly due to the difference of source properties between 
each tile and the intrinsic surface density fluctuations 
(e.g., between actual imaging region and white margin) within the tiles.

\begin{table}
\caption{Results of 7 different sets of simulations. \lq\lq z\rq\rq, \lq\lq s\rq\rq and
\lq\lq e\rq\rq indicate the redshift , size and ellipticity 
of the background sources, while \lq\lq 1\rq\rq and \lq\lq 0\rq\rq represent
following COSMOS and mocked. For example, \lq\lq e0\rq\rq 
implies the ellipticity (1-b/a) is randomly chosen in the range between 0 to 0.5. 
We run 10 lensing simulations 
in each projection of the 5 (middle) massive clusters at $z_{\rm l}=0.2$
for each source tile. In each case, 100 tiles are selected to construct the
control sample with required source properties. The redshift, ellipticity and size 
distributions of the selected COSMOS tiles are plotted 
in Fig.~\ref{fig:zdist}, \ref{fig:lwdist} and \ref{fig:sdist}, labeled as ``Control".
There are 15000 ray-tracing realizations performed for each control sample. (Similar results are obtained for clusters at $z_{\rm l}=0.3$).}

\begin{center}
\begin{tabular}{lclllll} 
\hline
Case & Name   &  z & s & e & N \\ \hline
1 & z1s1e1 & COSMOS & COSMOS & COSMOS & 1204\\
2 & z0s1e1 & z=1.0  & COSMOS & COSMOS & 1264  \\
3 & z0s0e0 & z=1.0 & $\Deff=1.0''$ & RANDOM & 321  \\
4 & z1s0e1 & COSMOS & $\Deff=1.0''$ &COSMOS& 852  \\
5 & z1s0e0 & COSMOS & $\Deff=1.0''$ & RANDOM & 369 \\
6 & z1s1e0 & COSMOS & COSMOS&  RANDOM & 493 \\
7 & z0s1e1 & z=1.5 & COSMOS & COSMOS & 1935 \\
\hline
\end{tabular}
\end{center}
\label{table:cases}
\end{table}

%%%% f9=com_steps.eps%%%%%%
%%%%%%%%%%%%%%%%%%%%%%%%%%%%%%
%%%%%%%%%%%%%%%%%%%%%%%%%

\begin{figure}
{
\epsscale{1.0}
\plotone{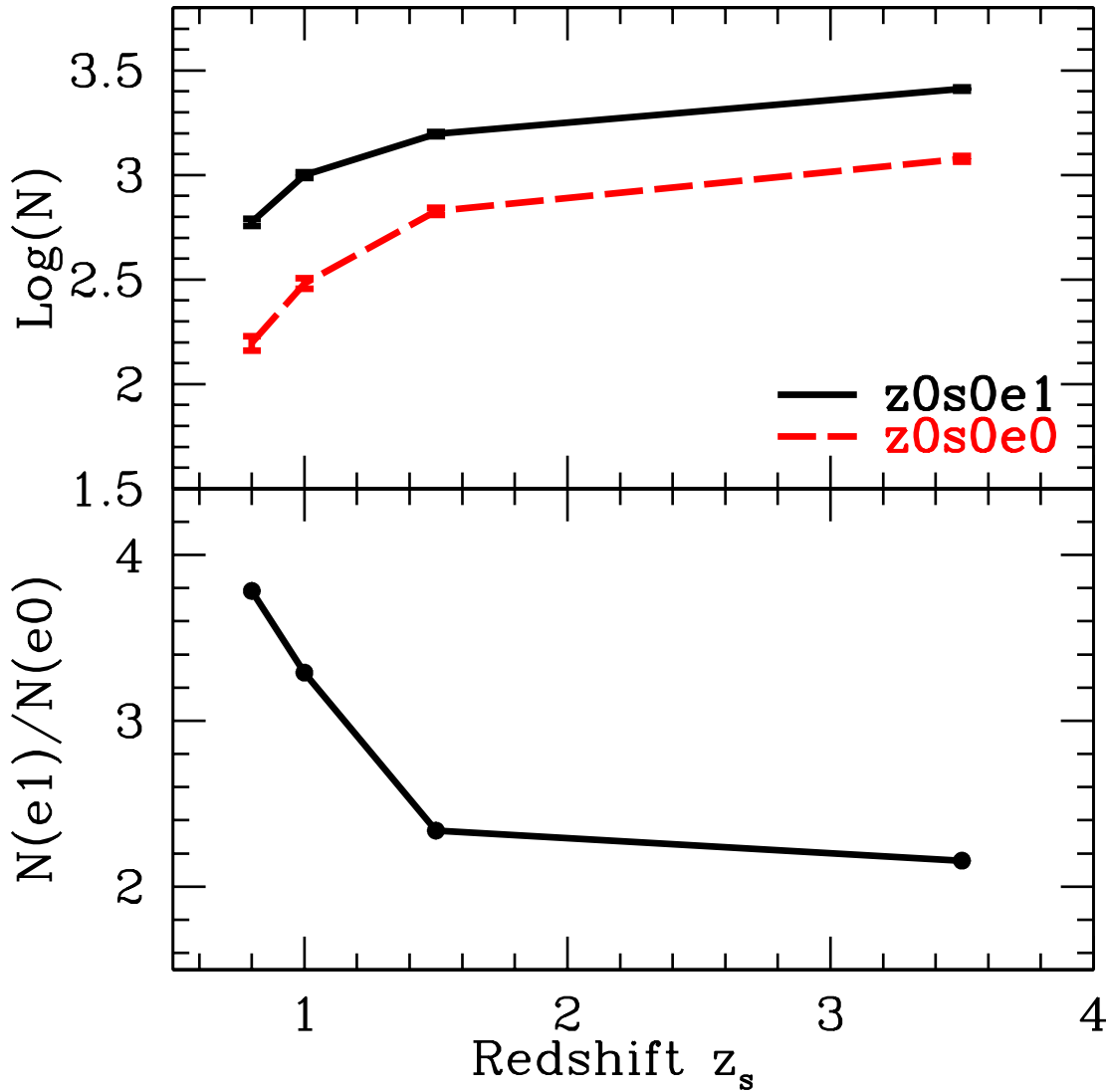} 
}
\caption{The shape impact on cross-section dependence of $z_{\rm s}$ is illustrated.
It is clearly seen that the lensing incidence is boosted after
the real galaxy shape is adopted (top panel).  
The Poisson fluctuations are indicated 
as error bars in the panel.  As a comparison, 
we put the sources of random ellipticity (shown in red) or 
real ellipticity (shown in black) in source planes at redshift
$z=0.8$, $z=1.0$, $z=1.5$ and $z=3.5$, keeping source size of $D_{\rm eff}=1''$.
While confirming the shape impact on lensing cross-section is about a factor of 2 for  
$z_{\rm s}>1.5$, with a larger factor at $z \la 1$ (bottom panel).
}
\label{fig:ze}
\end{figure}

\begin{table}
\caption{Ray-tracing simulations are carried out for the massive
	  clusters selected from two snapshots at redshift $\sim
	  0.2$ and $\sim0.3$. The cosmological model
          is a $\Lambda$CDM model with $\Omega_{\rm m,0}=0.268$,
$\Omega_{\Lambda,0}=0.732$, $\rm n=1$, and $\sigma_8=0.85$.
We implement 10 lensing simulations (marked as $N_R$) for each projection 
and each source tile.  
The whole simulation covers the full sample of 575 source tiles (marked as $N_T$).}
\begin{center}
\begin{tabular}{lcccccc} 
\hline
Set & $z_{\rm{l}}$& $M_{\rm vir}(10^{14}h^{-1}M_{\odot})$&$N_{\rm C}$ & $N_{\rm{P}}$& $N_R$ & $N_T$ \\ \hline
 1 & 0.2&6.8-11 & 10 & 3 & 10 & 575\\
 2 & 0.3&6.0-8.1 & 10 & 3  & 10 & 575\\
%6513  & 0.5 &4.7-6.2 & 5 & 3 \\
\hline
\end{tabular}
\end{center}
\label{table:simu}
\end{table}

\subsubsection{Redshift influence}

In recent works (\citealt{B98, Li05}), the background
sources are taken to be at the same redshift, e.g.,
$z=1$, to estimate the theoretical 
overall efficiency of forming giant arcs in our universe.
% which causes uncertainties on some level in the precise strong lensing
%statistics. 
As shown in Fig.~\ref{fig:zdist} the background sources in fact have a
rather broad distribution, which we take into account by binning them into $\sim 50$
redshift slices. If we put all the sources into 
a single redshift plane at $z_{\rm s}=1$, we find that in this case
the number of giant arcs ($L/W\geq 10$) increases by a factor of $\sim 1.05$ 
compared with Case 1 in Table~\ref{table:cases}, since the 
equivalent source redshift plane,
in which we put all the source galaxies to generate similar number of giant arcs,
for the 100 randomly selected COSMOS tiles 
	  (i.e., ``fiducial control sample" tiles which are considered as the ``seeds"
to generate mock samples for cases in Table~ \ref{table:cases}) is about 0.9
which is slightly lower than $z_{\rm s}=1.0$. %This shows that it is a relatively good approximation to put all the source at $z=1$ in our case.  
Notice that the shape, size distributions and surface density 
follow the original distributions of COSMOS.
Therefore, $z_{\rm s}$ plane is coincidently a good approximation to 
the broad redshift distribution.
However, according to \citet{Wambsganss04} and \citet{Li05}, 
the lensing probability becomes
larger when the sources are put at a higher redshift,  
which is also seen in the redshift distribution of lensed sources in
Fig.~\ref{fig:zdist} where the lensed galaxies, stretched to become giant arcs of $L/W>10$,
are clearly biased toward high
redshift relative to the underlying background population.
For comparison, the analytical redshift distribution in
\citet{Fedeli08} is also shown in Fig.~\ref{fig:zdist}.
If we adopt a source
redshift distribution following a deeper galaxy survey like in
\citet{Fedeli08} (keeping other properties unchanged), 
the giant-arc number is increased by a factor $\sim2$, 
compared with the number of giant arcs when the
COSMOS redshift distribution is used (Case 1 in Table~\ref{table:cases}). 

However, the equivalent source redshift plane changes with source shape 
	  distributions.
For example, considering the comparison cases, Cases 3 and 5 in Table~\ref{table:cases},
we find a higher equivalent source redshift (e.g., $z_s\sim 1.6$) for galaxies of simulated shapes as in Case 3 
than those of the intrinsic shape distribution in Case 1, which share the same intrinsic
redshift distribution. This can also be explained by Fig.~\ref{fig:ze}.
Galaxies of simulated ellipticities at higher redshift have a 
relatively higher weight than those of intrinsic ellipticities in producing giant arcs.
It causes the equivalent source redshift for Case 5 to be larger than the mocked redshift
in Case 3, e.g. $z_{\rm s}=1.0$, thus a higher lensing efficiency. 
Also, the redshift of equivalent source plane 
may shift to a higher redshift for a deeper survey. In this case, the lensing
probability may increase substantially and a single source plane
approximation (at $z_{\rm s}=1.0$) can become violated. For example, we find
an increase of a factor $\sim 2$ in giant-arc number for a mocked source sample 
following the analytical redshift distribution in \citet{Fedeli08}, 
which is shown in Fig.~\ref{fig:zdist}. 
It is necessary to take account of the full redshift distribution in predicting the 
strong lensing efficiency.

\subsubsection{Source shape influence}{\label{sec:sourceshape}}

Previous studies \citep{Li05, Puchwein09} often assumed the sources follow
a uniform distribution in the ellipticity ($\epsilon\equiv 1-b/a$) between
0 and 0.5. However,  as can be seen in Fig.~\ref{fig:lwdist}, while it is true that the
ellipticity $1-b/a$ of most sources is less than
0.5, there is still a high fraction of sources, about 1/3, with
ellipticity larger than 0.5. The sources with large ellipticities,
such as edge-on spiral galaxies, could be stretched much more easily to
form giant arcs of large $L/W$ values when the tangential direction of the
lens is approximately along the direction of the major axis of the
background galaxy. Therefore, the small portion of high ellipticity sources may
change the giant-arc statistics substantially.

For all the COSMOS sources in the fiducial control tiles, 
we change the shape of sources with
random ellipticities (between 0 and 0.5)
and orientations (between 0 and $\pi$) while keeping the 
source positions, sizes and redshifts unchanged. We perform a lensing simulation for
this background galaxy sample. This case (Case 6) is labeled as
``z1s1e0''. We find that the number of giant arcs is only $\sim 1/2$ of the value
from the COSMOS sources. The shape influence is also directly reflected in the different
ellipticity distributions of the original COSMOS
galaxies and the lensed population. As clearly seen 
in Fig.~\ref{fig:lwdist}, the lensed population 
of original galaxies with giant arcs of $L/W>10$ shows a clear shift to the higher ellipticity tail.

It is important to explore whether the effect of source shape changes with the 
source redshift. Although the overall cross-section becomes larger for a higher source
redshift, the change of the effective lensing area, 
in which elongated sources can be lensed to giant arcs while 
the simulated ones of $0\leq\epsilon\leq0.5$ cannot be, is not clear.
We construct 4 comparison pairs of control source samples generated from the fiducial
control tiles, 
i.e. ``z0s0e1'' and ``z0s0e0'', at redshift of $z_{\rm s}=0.8, 1.0, 1.5, 3.5$.
As clearly seen in Fig.~\ref{fig:ze}, 
while the number of giant arcs increases with the source redshift, the shape impact on 
lensing efficiency decreases from $\sim 4$ to $\sim 2$, when source redshift ranges
	  from $z_{\rm s}$=0.8 to 3.5.
The boost effect of source shape is expected to be 
larger at $z_{\rm s}$ lower than 1.5 since the lensing cross-section is smaller at lower redshift and thus the enhancement due to the elliptical sources is relatively more important.
%although the fluctuation is bigger at lower source redshift due to 
%the dramatic number decrease of giant arcs.

\subsubsection{Source size influence} {\label{sec:sourcesize}}

Instead of assuming a same effective diameter for all sources
\citep{Li05, Puchwein09}, we have adopted the 
COSMOS galaxy sizes in our fiducial case (Case 1).
%with real extracted images whose effective diameters range from $0.5''$ to $1.5''$. 
To see the size effect of background galaxies, we first check the size 
distribution for the lensed galaxies which are stretched to be giant arcs of $L/W>10$. 
The comparison results are illustrated
in Fig.~\ref{fig:sdist}, it shows that there is little difference between 
these two distributions,
and thus the source size is expected to have small effect on producing giant arcs.

\begin{figure}
{
\epsscale{1.0}
\plotone{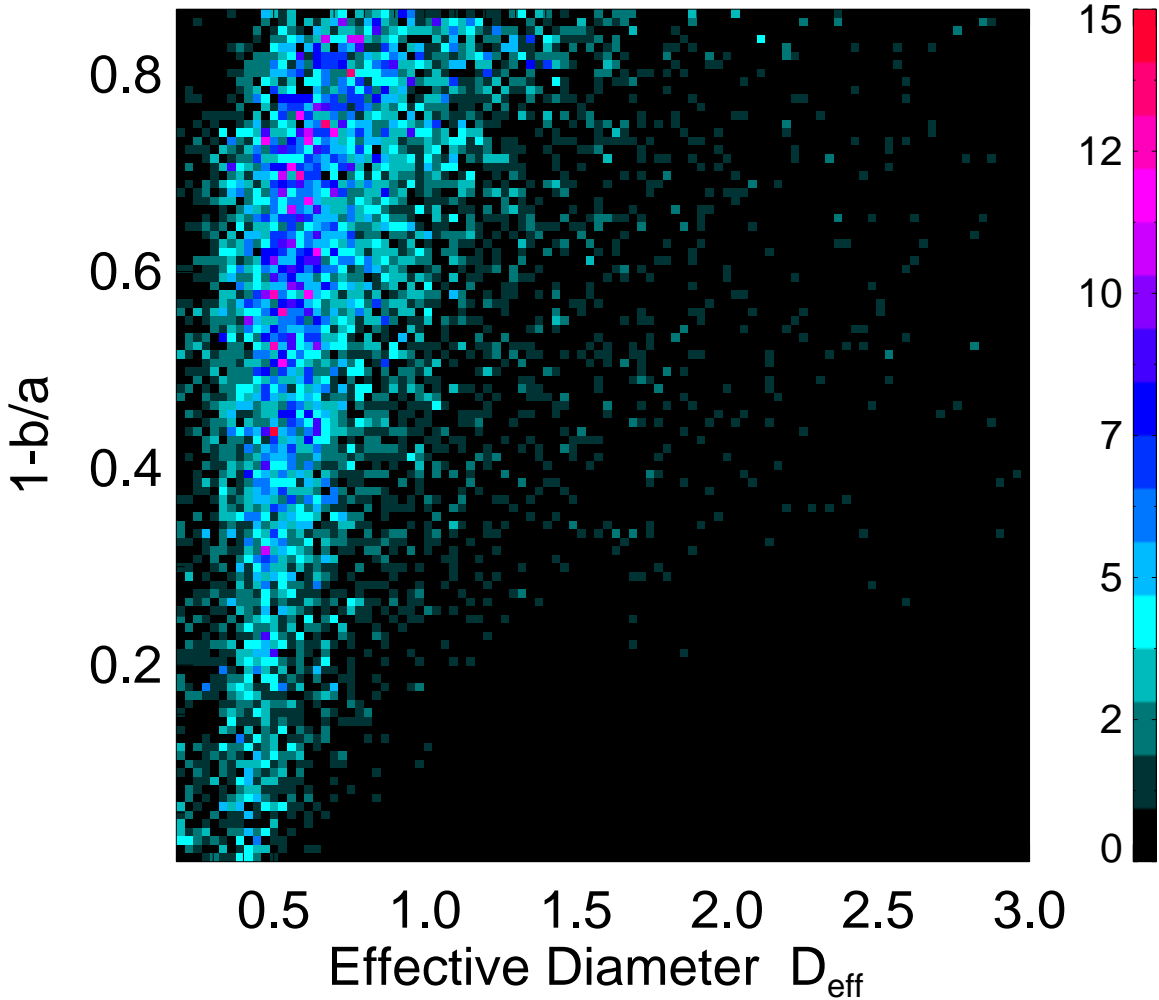}
}
\caption{The color-weighted distribution plot of the ellipticity against size for 
sources lensed to be images with $L/W\geq 10$.
  The lensed sample is obtained from tracing back all the giant arcs produced
  by 10 massive clusters at redshift $z \sim 0.2$.
  The pattern of the distribution is
  similar to that shown in Fig.~\ref{fig:slw}, although the elongated sources have
  a higher weight, which is consistent with that implied in Fig.~\ref{fig:lwdist}.
%Then, their ellipticity distribution is plotted against the source area size.
} 
\label{fig:lwsmap}
\end{figure}

This is directly confirmed by two comparison samples, where we fix the effective diameters 
assigned to be $\Deff=1.0''$ and $\Deff=0.5''$.
We find a similar number of giant arcs (188 and 166 out of 3000 realizations), 
which is consistent with Fig.~\ref{fig:sdist}. 
%and $109$ in Case 6 (\lq\lq z1s0e1\rq\rq).  Fig.~\ref{fig:sdist} and
%Fig.~{\ref{fig:ca02}} show the comparison results. 
Source size thus has a moderate effect on producing giant arcs 
when massive clusters are taken as lenses 
(also see Fig.~\ref{fig:slw} and Fig.~\ref{fig:lwsmap}). The reason is that the
typical source size is much smaller than the 
strong lensing caustic size, and thus a moderate change in the source
size has a small impact on the number of giant arcs.
%It is probably more 
%sensitive for the source position in the caustic area rather than the size of the galaxy.

%%%% f11=lwsmap.eps%%%%%%
%%%%%%%%%%%%%%%%%%%%%%%%%%%%%

\subsection{Combination of Morphology and Redshift Distribution}

To see the overall impact of the simple assumptions for the shape,
size, and redshift of background galaxies,  we combine these
effects to do lensing simulations (for random position
effect, see \S\ref{sec:clustering}).  The redshift of all these
sources
% (\lq\lq z0s0e0\rq\rq in Fig.~\ref{fig:ca02})
is assigned to be 1.0. The shape with a random orientation 
(between 0 and $\pi$) are simulated as above, while the 
effective diameter is assigned to be $1''$.
We find that the lensing efficiency for the Case 3 (``z0s0e0'')
is a factor of 3.75 lower than that for Case 1 (``z1s1e1''). It is roughly
consistent with the results from the combination of the individual effects, i.e. redshift effect 
(a factor of 1.15 which is got from comparing Case 3 and 5), shape
effect (a factor of 2.31 from Case 4 and 5) and size effect (a factor of 1.34 from Case 5 and 6),
which is $1.15\times2.31\times1.34=3.56$. 
It thus indicates that the galaxy shape, 
redshift and size are affecting strong lensing efficiency independently as well as these
three properties are independent of each other intrinsically as shown in Fig.~\ref{fig:sz}, \ref{fig:lwz} and \ref{fig:slw}.
%Notice that the morphology impact of background galaxies will vary with
%the source redshift distribution, 
%which could be seen by comparing the number ratio of Case 1 
%versus Case5 and Case 2 versus Case 3. The morphology dependence becomes smaller when the
%source redshift is increased.

\subsection{Effect of Source Clustering}{\label{sec:clustering}}

Theoretical calculations often assume that the background sources
are randomly distributed in space. However, 
background galaxies are clearly clustered. It is therefore interesting to check
the effect of source clustering on the incidence of giant arcs.

The influence of the clustering effect is investigated by carrying out a
ray-tracing simulation on a control sample of galaxies where
the positions of original COSMOS galaxies in a fiducial tile are randomly shuffled while
keeping the source surface density, size and ellipticity
unchanged. We find that the efficiency of producing giant
arcs (166 giant arcs out of 3000 realizations for 10 clusters at $z_{\rm l}=0.3$) is similar to what obtained in the COSMOS case (163 giant arcs) which implies the
clustering effect of background sources is negligible on strong
lensing. The likely reason is that the intrinsic source density is too low (43 per
square arcmin)
and close pairs of sources are too few to form connected
giant arcs or to be strongly lensed simultaneously. 
Our results suggest that sources can be assumed to be random
in the prediction of giant arcs, at least to the depth of COSMOS (in
deeper images with a larger source density, clustering may become more important).

\subsection{Effects of Seeing in Ground-based Observations}{\label{sec:seeing}}

For our fiducial simulation (Case 1), the observational data are taken
by HST (as in the Hubble Deep Field used in Horesh et al. 2005). However, many giant
arcs were discovered from the ground-based telescopes 
where the effect of seeing may be important.
A proper evaluation would require a detailed simulation of seeing effects.
Here we discuss briefly the approximate seeing effect on the predicted
number of giant arcs. 

In COSMOS, there is a significant portion 
($\sim80\%$) of small galaxies with the effective diameter $D_{\rm eff}\leq
1''$ (see Fig.~\ref{fig:sdist}). The width of lensed giant arcs produced from
these small sources probably remains less than $1''$.  In such cases, seeing (such as $\sim1''$) will
have a much larger effect on the width of giant arcs than on the length
\citep{Mene07}, and in general the $L/W$ ratio will be substantially reduced.
Assume the seeing is $\sim 0.5''$, the width of arcs intrinsically smaller
than $0.5''$ will be blurred to $\sim 0.5''$ which we adopt as the
``observed'' width. We find that in this case the fraction of giant arcs of ${L/W}\geq10$ will be reduced to
$80\%$ of the COSMOS value. If seeing is as bad as $1''$, the lensing probability
could be reduced to about $40\%$ of the original value.
Although this is only an approximate estimation, it is clear that the impact of seeing should be
carefully evaluated in a detailed comparison between theoretical
predictions and ground-based giant-arc surveys.

\subsection{Comparison with Observations and Horesh et al. (2005)} {\label{sec:le}}

The giant arcs in this work are selected by quantifying the ratio of length
to width ($L/W \geq10$). The definition of L/W in this paper is
the same as that in \citet{Li05}. As a mock strong lensing survey, we have done 
172500 realizations of ray-tracing simulations for all the
10 lens clusters at $z_{\rm l}=0.2$ or $z_{\rm l}=0.3$ with 575 galaxy tiles, and find that 
the lensing efficiency is about $\sim0.0743$ arcs for 
clusters at $z_{\rm l}=0.2$ and $\sim 0.0563$ arcs for clusters at $z_{\rm l}=0.3$ per realization. 

%giant arcs, which makes the length to width ratio a factor of
%$4/\rm\pi$ smaller than the rectangular fitting in \citet{Horesh05}.

On the face value, they are much lower than the
expected mean probability of $\sim 1$ giant arc per realization
(\citealt{Horesh05}). However, they adopted galaxy images in the Hubble 
Deep Field whose source density 
is about 12 times higher (1378/5.3=260 ${\rm arcmin}^{-2}$ in HDF, vs. 23 
		    $\rm {arcmin}^{-2}$ in our case).
The lensing efficiency with the surface density normalized is 0.840 arcs and 0.636 per realization, respectively.
 \citet{Horesh05} used a different definition of $L/W$ ratio
which is smaller than ours by a factor of $4/\pi$ (e.g., their $L/W\geq10$
is equivalent to our $L/W\geq2.5 \pi \approx 7.5$), and thus our lensing efficiency 
(of $L/W\geq10$) will 
be boosted by a factor of $(4/\pi)^2$ if we adopt their length-to-width ratio definition.
Taken these two factors into account, our clusters produce $0.840 \times
(4/\pi)^2=1.36$ and $0.636 \times (4/\pi)^2=1.03$
giant arcs per realization for $z_{\rm l}=0.2$ and 0.3, respectively. In this exercise, we 
have assumed that the giant arc formation cross-section scales
as the magnification probability distribution ($p(>|\mu|) \propto \mu^{-2}, |\mu| \gg 1$). 
Therefore, the capability of our simulated lens clusters is 
in fact consistent with the
observation, which is about 1.2 arcs per cluster found by \citet{Horesh05} using
the sample in \citet{Smith05}.

The slight excess in the number counting could be induced by 
small number statistics and the relatively narrow mass range of the simulated lenses by
contrast to the observational clusters. 
It also could be due to the different lens redshift distributions of the simulated 
($z_{\rm l}\sim0.2$) and observational samples ($0.171<z_{\rm c}<0.255$).

%the bigger simulation box and thus higher mass range
%(for example, the lensing efficiency is about 0.114 for the first 5 massive lenses 
% and 0.0347 for the 5 less massive clusters) and
%partly due to the fluctuation because of a smaller simulated cluster sample in \citep{Horesh05}.

%The statistics of giant arcs in 3 runs of 5 massive clusters for 20
%tiles are illustrated in the Fig.~\ref{fig:com_steps}.

%\input{dis_ack}
\section{Summary and Discussion}

In this paper, aiming to study the impacts of background sources on strong lensing
statistics, we use the $I$-band galaxy image data of HST/ACS in the COSMOS to
quantify the distributions of background source size, shape and
redshift. Each galaxy image is extracted by SExtractor
limiting a surface brightness down to $\sim 25 \,{\rm mag/arcsec^2}$.  
The redshift of each
galaxy image is obtained by matching its celestial position with the COSMOS
photometric catalog. The selected sources are then lensed by
10 massive clusters of the mass range
$6.8 \times 10^{14} h^{-1} M_{\odot}\leq\rm M_{\rm vir}\leq1.1\times
10^{15} h^{-1} M_{\odot}$ at $z_{\rm l} \sim 0.2$ and 
$6.0 \times 10^{14} h^{-1} M_{\odot}\leq\rm M_{\rm vir}\leq 8.1\times 
10^{14} h^{-1} M_{\odot}$ at $z_{\rm l} \sim 0.3$ as the lensing
clusters chosen from a cosmological simulation in the $\Lambda$CDM
cosmology ($\Omega_{\rm m,0}=0.268, \Omega_{\Lambda,0}=0.732$).
575 source tiles within the COSMOS field of around 2 ${\rm deg^2}$ 
are fully used for the statistic study. 
172500 ray-tracing lensing simulations are carried out for 10 lens clusters to
reduce the statistical fluctuation at the expected lens redshifts (i.e. $z_{\rm l}=0.2$ and 0.3).
The incidence of giant-arc production in our simulation is 
roughly consistent with that observed in \citet{Smith05},
after the density difference is taken into account.
The impacts of source size, shape and
redshift on strong lensing statistics are investigated in detail.

We find that the source size (less than a factor of 1.5) and clustering only have small
effects on the production of giant arcs. In contrast, the dependencies on the
source redshift and ellipticity are much more significant. The first was
highlighted by \citet{Wambsganss04}, while the second is the main new finding of
the current work. We find that adopting the empirical ellipticity
distribution of COSMOS increases the lensing probability by a factor of
2 (see Fig.~\ref{fig:ze} and also the number ratio between Case 1 and 6).
The boosting effect of the
ellipticity of the background galaxies has not been emphasized in the previous works. It
should be included in theoretical modeling of giant arcs in future.
This may also be of particular relevance for setting constraints on the
power-spectrum normalization parameter $\sigma_8$.

There are a number of limitations in the present work.
When scaling our results with the source number density, we implicitly assume that 
other source properties (like ellipticity and redshift) are the same in COSMOS as in a
deeper observation (like HDF). However, ellipticity distribution of galaxies in a deeper survey
may turn out to be more elliptical (\citealt{Vincent05}; see also the
discussion on the surface 
brightness limit), which would increase the boosting factor of source ellipticity on strong lensing efficiency. We 
would also have a larger fraction of galaxies at the high redshift tail end in a galaxy survey 
like HDF, which increases the mean strong lensing cross section. 
Therefore, both effects will  produce more giant arcs.
For direct comparison with observation, the observational
effects, such as specific instrumental point-spread-function and observational seeing, are
not fully included in our lensing simulations. 
	  As mentioned in \S\ref{sec:seeing}, if the seeing was
$1''$, then there may be only around $40\%$ of giant arcs observed by a typical
ground-based telescope compared with that by space-based one, such as HST.

The same shape measurement method is used for quantifying the ellipticity of the
original galaxies and lensed images. It could bring in some inaccuracy
for measuring the ellipticity of very round images, 
since it is mainly designed for quantifying arcs. Nevertheless, this effect is rather 
small and negligible to our results, since we mainly focus on the elliptical galaxies. 
Besides, the intrinsic pixelization will
affect the ellipticity quantification of small sources, especially for sources of 
$D_{\rm eff}<0.3''$ (see Fig.~\ref{fig:slw}). The influence of pixelization would 
probably shift down the measured ellipticity because of 
its relative larger effect on width measurement.  Since the number of such tiny sources is 
small ($\sim 5\%$) and the pixelization does not significantly change the ellipticity
value on average (much smaller than the effect of seeing), 
this influence would not be important to our analysis.

According to \cite{Torri04}, cluster substructures are important for strong lensing 
efficiency in a major merger case, especially at the early state. 
As we adopt a softening length of 30 $h^{-1}\rm{kpc}$, 
the softening mainly takes effect in the center of 
major merging clumps, while it would also smooth away the small substructures of 
a similar scale in a minor merger case or those remaining in a main halo. 
The change of the lensing cross section due to this kind of smoothing would
not be significant in both cases. Besides, according to a comparison plot of optical depth 
in Fig.~4 
of \cite{Li05}, the result agrees within $25\%$ of \cite{Wambsganss04}, in which they 
adopt a much smaller softening length of $3.2 h^{-1}\rm{kpc}$ than 30 $h^{-1}\rm{kpc}$ in
our case, for a source redshift of $z_{\rm s}=1.0$. Therefore, although 
the relatively large value of softening length could reduce the lensing efficiency, it 
would not affect our results significantly.
%Although we can estimate the effect of seeing by a simple assumption,
%PSF effect is difficult to model both in simulation and observation
%and varies with different observational facilities.  
Moreover, lensing clusters are 
selected from a dark matter only simulation, therefore, we cannot
investigate the strong lensing dependence on source properties for
clusters with realistic baryon distributions.

%The dependence could be weaker compared with the
%lensing by clusters in $N$-body simulations (why???), because of the increase of lensing
%probability with about a factor $\sim 2$ \citep{Mene03a}.  As an
%indication from the results obtained by using the redshift
%distribution of \citet{Fedeli08}, from this angle of view, a deeper
%survey is also promising to solve the problem proposed by\citet{B98}. 
We have only used one surface brightness limit (e.g., a detection threshold of 1.5
$\sigma$) to extract the background galaxy images.
Since the redshift of galaxy is mainly constrained
by the matching with photometric catalog, it 
will not change the distribution by using a higher detection threshold in our case.
A variation of detection threshold would change the galaxy size, 
  but the dependence of lensing efficiency on source size is weak 
  (see \S\ref{sec:sourcesize}), thus the main results should remain unchanged.
The source shape distribution also depends on the surface brightness threshold.
Indeed, at a lower surface brightness limit, the shape of the galaxies tends to more 
elliptical \citep{Vincent05}. Thus at fainter surface brightness,
the shape influence is expected to play a more important role on giant-arc 
production for the same lens cluster population.
It is meaningful to do ray-tracing experiments by using several surface brightness limits
to quantify the shape impact on strong lensing in different survey depths.

The different mass range of the simulated clusters and X-ray selected ones 
could bring in quantitative uncertainties in the giant-arc statistics, since the 
strong lensing cross section strongly depend on mass, e.g., the mean cross section changes
by nearly 4 orders of magnitude over mass ranging from $10^{14}$ $h^{-1}{M_{\odot}}$ to 
$\sim10^{15}$ $h^{-1}{M_{\odot}}$ in Fig.~7 of \cite{Hen07a}. 
However, the mean cross section ratio is less than 1.5 between the observed and our simulated clusters, according to the plot. 
Moreover, they use a larger mass density parameter $\omega0=0.3$ and a higher 
matter power-spectrum normalization $\sigma_8=0.95$ in their simulation. Besides, the 
simulated clusters are put at a much  higher redshift $z_{\rm l}=0.41$. All these effects 
help to make the mass dependence of lensing probability stronger. 
Therefore, the difference of mean cross section
in two comparing samples would be much less than a factor of 1.5 
and becomes negligible in this work.
Nevertheless, it is still necessary and interesting to select more comparable 
(between the simulated and X-ray selected) lens cluster samples in mass and redshift range
for a fair comparison of giant-arc producing efficiency in the future.

\section*{Acknowledgment}

We would like to thank the anonymous referee for his/her useful suggestions. 
We also thank W.~P. Lin for managing the large amount 
of the data and helpful discussions. 
This work is supported by NSFC (10533030, 10821302, 10878001), by the
Knowledge Innovation Program of CAS (No. KJCX2-YW-T05), and by 973
Program (No. 2007CB815402).  GL is supported by the Humboldt
Foundation.  SM acknowledges the Chinese Academy of Sciences and the
Chinese National Science Foundation for travel support. This work was
also partly supported by the visitor's grant at Jodrell Bank, a joint
research grant from the NSFC and the Royal Society,
the Department of Energy contract DE-AC02-76SF00515 and by the European
Community's Sixth Framework Marie Curie Research Training Network
Programme, Contract No. MRTN-CT-2004-505183 ``ANGLES''.

\end{document}